\documentclass[aip,jcp,11pt,reprint,floatfix,superscriptaddress,urlname]{revtex4-1}
\usepackage[utf8]{inputenc}
\usepackage{amsmath}
\usepackage{color}
\usepackage{graphicx}
\usepackage{comment}
\usepackage[normalem]{ulem}

\newcommand{\ket}[1]{\left| #1 \right>} % for Dirac bras
\newcommand{\bra}[1]{\left< #1 \right|} % for Dirac kets
\newcommand{\matrixel}[3]{\left< #1 \vphantom{#2#3} \right|
 #2 \left| #3 \vphantom{#1#2} \right>} % for Dirac matrix elements

\begin{document}
\title{Conservation laws in coupled cluster dynamics at finite-temperature}

\author{Ruojing Peng}
\affiliation{Division of Chemistry and Chemical Engineering, California Institute of Technology, Pasadena,
California 91125, USA}
\thanks{R. Peng and A. F. White contributed equally to this work}

\author{Alec F. White}
\affiliation{Division of Chemistry and Chemical Engineering, California Institute of Technology, Pasadena,
California 91125, USA}
\thanks{R. Peng and A. F. White contributed equally to this work}

\author{Huanchen Zhai}
\affiliation{Division of Chemistry and Chemical Engineering, California Institute of Technology, Pasadena,
California 91125, USA}

\author{Garnet Kin-Lic Chan}
\affiliation{Division of Chemistry and Chemical Engineering, California Institute of Technology, Pasadena,
California 91125, USA}

\begin{abstract}
    We extend the finite-temperature Keldysh non-equilibrium coupled cluster theory (Keldysh-CC) [{\it J. Chem. Theory Comput.} \textbf{2019}, 15, 6137-6253] to include a time-dependent orbital basis. When chosen to minimize the action, such a basis restores local and global conservation laws (Ehrenfest's theorem) for all one-particle properties, while remaining energy conserving for time-independent Hamiltonians. We present the time-dependent orbital-optimized coupled cluster doubles method (Keldysh-OCCD) in analogy with the formalism for zero-temperature dynamics, extended to finite temperatures through the time-dependent action on the Keldysh contour. To demonstrate the conservation property and understand the numerical performance of the method, we apply it to several problems of non-equilibrium finite-temperature dynamics: a 1D Hubbard model with a time-dependent Peierls phase,
    laser driving of molecular H$_2$, driven dynamics in warm-dense silicon, and transport in the single impurity Anderson model.
    %to demonstrate particle number conservation of Keldysh-OCCD method over the previous result. The Keldysh-OCCD method is also applied to a molecular laser-driven H$_2$ system and the single impurity Anderson model(SIAM) with a quenched Hamiltonian. 
    %    For the SIAM model, we compute the time-dependent current through the impurity, from which we extract the conductance behaviour to compare against other numerical methods, such as finite-temperature density matrix renormalization group.
\end{abstract}

\maketitle

\section{Introduction}

The simulation of real-time electronic dynamics in molecules and materials is a challenge due to the many-body nature of the Hamiltonian combined with the need to propagate for many time steps in order to see the relevant phenomena. The problem is further complicated when finite-temperature effects are important, as is the case in many condensed-phase systems. 

In {\it ab initio} electron dynamics, a variety of methods have been applied to simulate various phenomena, such as multiple photon processes \cite{Huang1994}, high harmonic generation \cite{Sato2016}, and ultrafast laser dynamics \cite{Huber2011} in atomic and molecular systems. Among them are time-dependent density functional theory\cite{Runge1984} and wavefunction based methods such as time-dependent Hartree-Fock\cite{Kulander1987}, configuration interaction (CI)\cite{Klamroth2003,Krause2005,Schlegel2007,Krause2007}, complete-active-space self-consistent field (CASSCF)\cite{Sato2016}, multiconfigurational time-dependent Hartree-Fock\cite{Kato2004,Nest2005,Kato2008}, density matrix embedding theory (DMET)\cite{Kretchmer2018} and coupled cluster (CC) methods \cite{Schonhammer1978,Hoodbhoy1978,Hoodbhoy1979,Sato2018CC,Pederson2021}.  \textit{Ab initio} electron dynamics in materials has primarily been carried out at the time-dependent density functional theory level, although studies with low-order diagrammatic approximations have begun to appear\cite{attaccalite2011real}.
In more correlated materials, there are many studies involving model Hamiltonians, for which methods such as density matrix renormalization group (DMRG)\cite{Cazalilla2002,Verstraete2004,Kokalj2009,Wolf2014,Ren2018} and diagrammatic Monte Carlo\cite{Werner2009,Schiro2009,Segal2010,Werner2010,Antipov2017} are popular.

Here, we are interested in  the dynamics of {\it ab initio} electronic Hamiltonians at finite-temperature in both condensed-phase and molecular systems. We include finite-temperature from the outset to address certain classes of problems. For instance, thermal effects are
already important at the spin exchange energy scale in correlated materials with low-temperature electronic phase transitions~\cite{Vojta2000,Lee2006}. Temperature also plays a role in the coupling of electrons with lattice vibrations\cite{Cronin2006,Waldecker2016,Wright2016}. Finally, the study of matter under extreme conditions such as those found in planetary cores\cite{Fortov2009} also necessitates finite-temperature methods. 

Various wavefunction methods traditionally used for \textit{ab initio} zero-temperature quantum chemistry have been extended into the finite temperature regime. For example, the equilibrium finite temperature coupled cluster equations have been previously derived via the thermal cluster cumulant (TCC) method\cite{Sanyal1992,Sanyal1993,Mandal2001,Mandal2002} and more recently through a time-dependent diagrammatic approach\cite{White2018,White2020}. 
 CI \cite{Harsha2019CI} and CC \cite{Harsha2019CC,Shushkov2019} methods have also been extended into the finite temperature regime by means of the thermofield dynamics language. 
In a similar manner, there
are also several formalisms to generalize equilibrium finite temperature theories to non-equilibrium dynamics. Within a time-dependent diagrammatic language, non-equilibrium dynamics corresponds to moving time integration from the imaginary axis to the Keldysh (or Kadanoff-Baym) contour in the complex time plane\cite{Keldysh1965,kadanoff2018quantum}. In the context of coupled cluster theory, this diagrammatic extension leads to the Keldysh coupled cluster (Keldysh-CC) theory~\cite{White2019}.
%that can describe not only finite-temperature equilibrium properties, but also their dynamics.
%of {\it ab initio} Hamiltonians at finite temperature. 

In traditional Feynman diagrammatic approximations, a set of sufficient conditions, known as the ``conserving" conditions~\cite{Baym1961,Baym1962}, are known. When diagrammatic approximations for the self-energy are constructed to satisfy these conditions, the resulting approximate single-particle Green's function dynamics satisfies the physical conservation laws that, for example, relate the time-derivative of local single-particle observables such as the density and momentum density to their corresponding currents.
However, the  Keldysh-CC method is not constructed as a conserving approximation, and this can lead to unphysical results when propagated for long times. It is this issue which we aim to correct. 

It is known from time-dependent wavefunction theories that the use of time-dependent orbitals that obey an equation of motion derived from a time-dependent variational principle (TDVP) \cite{Dirac1930,McLachlan1964,Broeckhove1988} leads to the conservation of 1-particle properties in the sense that Ehrenfest's theorem
\begin{equation}\label{ehrenfest}
    \frac{d}{dt}\langle A\rangle=-i\langle[A,H]\rangle+\langle\frac{\partial A}{\partial t}\rangle
\end{equation}
is satisfied for the 1-particle reduced density matrix (1RDM). At zero-temperature, this type of orbital dynamics has been used for time evolution with various \textit{ab initio} methods, including  time-dependent CI \cite{Sato2018CI}, CASSCF\cite{Sato2016}, DMET\cite{Kretchmer2018}, and CC\cite{Sato2018CC,Kvaal2012,Pedersen1999a,Pederson2021}. Starting from this same idea, we can extend the Keldysh-CC method to include the variational orbital dynamics via  a finite-temperature TDVP. This restores the consistency and local conservation laws for all 1-particle properties, and further maintains the global conservation law for the energy present in the original Keldysh-CC theory. 

In Section~\ref{sec:theory} we review ground-state and finite-temperature coupled cluster theory for systems in and out of equilibrium. We discuss some of the deficiencies of the finite-temperature, non-equilibrium Keldysh-CC theory presented in Ref.~\onlinecite{White2019} and we show how we can remedy some of these problems with orbital dynamics in analogy with ground state theories. This leads to the derivation of the Keldysh orbital-optimized coupled cluster doubles (Keldysh-OCCD) method. In Section~\ref{sec:result}, we apply the method to the model problem of a 2-site Hubbard model with a Peierls phase as well as an {\it ab initio} model of laser driven warm-dense silicon discussed in Ref.~\onlinecite{White2019}. In both of these, we demonstrate the exact conserving behaviour of the theory.
%and demonstrate some improvement of the Keldysh-OCCD method over Keldysh-CCSD. 
We then further assess the numerical behaviour of the method in an application to laser driven molecular H${}_2$, as well as non-equilibrium transport in the 1D single impurity Anderson model (SIAM).  %\textcolor{red}{comparing to exact dynamics in the former,} and 
The performance of Keldysh-OCCD on SIAM is compared to a numerically accurate benchmark result from finite-temperature DMRG.
%in the latter.

\section{Theory}\label{sec:theory}

\subsection{Coupled cluster dynamics at zero temperature}

The ground-state coupled cluster method is defined by an exponential wavefunction ansatz,
\begin{equation}
    \ket{\Psi_{\mathrm{CC}}} = e^T\ket{\Phi_0}
\end{equation}
where $\Phi_0$ is a reference Slater determinant. The $T$ operator is defined in a space of excitation operators indexed by $\nu$,
\begin{equation}
    T = \sum_{\nu}t_{\nu}\hat{\nu}^{\dagger},
\end{equation}
where $\hat{\nu}^{\dagger}$ excites the reference determinant such that
\begin{equation}
    \ket{\Phi_{\nu}} = \nu^{\dagger}\ket{\Phi_0}.
\end{equation}
This space of excitations is usually truncated based on excitation level. For example, constraining $T$ to the space of single and double excitations from the reference yields the commonly-used coupled cluster singles and doubles (CCSD) method. The coupled cluster energy and amplitudes are determined from a projected Schrodinger equation:
\begin{align}
    \matrixel{\Phi_0}{\bar{H}}{\Phi_0} &= E_{\mathrm{CC}} \\
    \matrixel{\Phi_{\nu}}{\bar{H}}{\Phi_0} &= 0. \label{eq:projections}
\end{align}
Here, we have written the similarity-transformed Hamiltonian as
\begin{equation}
    \bar{H} \equiv e^{-T}He^T.    
\end{equation}
Properties are computed from response theory which is, in practice, accomplished by solving for Lagrange multipliers, $\lambda^{\nu}$ associated with the solution conditions~\eqref{eq:projections}. These appear in the Lagrangian,
\begin{equation}
    \mathcal{L} = \matrixel{\Phi_0}{\bar{H}}{\Phi_0} + \sum_{\nu}\lambda^{\nu}\matrixel{\Phi_{\nu}}{\bar{H}}{\Phi_0}.
\end{equation}
The Lagrange multipliers may be associated with elements of a de-excitation operator,
\begin{equation}
    \Lambda = \sum_{\nu} \lambda^{\nu} \hat{\nu},
\end{equation}
such that $\bra{\Phi_0}(1 + \Lambda)$ is the left eigenstate of the projected Schrodinger equation.

The coupled cluster model can be extended to treat dynamics by allowing $T$ and $\Lambda$ to become functions of time. The appropriate equations of motion may be obtained from an action,
\begin{equation}\label{eqn:ccaction}
    S = \int_{t_i}^{t_f}d t' 
    \matrixel{\Phi_0}{(1 + \Lambda)e^{-T}(H - i\partial_t)e^T}{\Phi_0},
\end{equation}
by making it stationary with respect to variations in the $T$ and $\Lambda$ amplitudes. The resulting equations of motion are given by
\begin{align}
	i\dot{t}_{\nu} &= \matrixel{\Phi_0}{\nu \bar{H}}{\Phi_0} \label{eqn:EOM_T}\\
	-i\dot{\lambda}^{\nu} &= \matrixel{\Phi_0}{(1 + \Lambda)e^{-T}[H,\nu^{\dagger}]e^T}{\Phi_0}\label{eqn:EOM_L}.
\end{align}
Here $\nu^{\dagger}$ is an excitation operator and $\nu$ is the corresponding de-excitation operator. The properties of such coupled cluster dynamics have been discussed in many works over the years\cite{Arponen1983,Dalgaard1983,Koch1990,Pedersen1997,Pedersen1999b,Pedersen2001}. In particular, we note that conservation laws are not generally obeyed and Ehrenfest's theorem (Equation \ref{ehrenfest}) is not generally satisfied\cite{Pedersen1997,Pedersen1998,Pedersen1999a,Pedersen1999b}. However, there are some special cases in which Ehrenfest's theorem will be satisfied. Energy will be conserved for a time-independent Hamiltonian, and particle number will be conserved also. 

\subsection{Coupled cluster dynamics with time-dependent orbitals}\label{sec:zt-orb-dyn}
It has been shown that making the above action (Equation~\ref{eqn:ccaction}) stationary with respect to time-dependent orbitals  leads to dynamics that satisfy Ehrenfest's theorem for all 1-electron operators\cite{Sato2018CI,Sato2018CC,Sato2016,Pedersen1999a,Pederson2021}. This can be important because, as pointed out by Pedersen {\it et al}\cite{Pedersen1999a}, it leads to local conservation and gauge invariance in 1-electron properties.

Orbital dynamics can be included by adding a unitary, orbital-rotation operator to the wavefunction ansatz:
\begin{equation}
	\ket{R} \equiv e^{\kappa}e^{T}\ket{\Phi_0} \qquad \bra{L} \equiv \bra{\Phi_0}(1 + \Lambda)e^{-T}e^{-\kappa}.
\end{equation}
Here, $\kappa$ is an anti-Hermitian, time-dependent 1-electron operator. A Lagrangian of the form
\begin{equation}
    \mathcal{L} = \matrixel{R}{H - i\partial_t}{L}
\end{equation}
will lead to a biorthogonal approach termed OATDCC by Kvaal\cite{Kvaal2012}. Alternatively, one may use a Lagrangian given by
\begin{equation}
    \mathcal{L} = \frac{1}{2}\left[\matrixel{R}{H - i\partial_t}{L} + \mathrm{c.c.}\right]
\end{equation}
which has the advantage of being real and leads to orthonormal orbital equations. This is the approach taken by Pedersen {\it et al} and later by Sato {\it et al}\cite{Sato2018CC}. We take the analogous approach in this work.

To derive the equations of motion for such a theory, it is convenient to define a 1-electron operator,
\begin{equation}
    X \equiv e^{-\kappa}\partial_te^{\kappa}.
\end{equation}
Up to this point, we have assumed a fixed set of reference orbitals, but it is easier to represent the equations using a picture where the orbitals of the reference, and corresponding representation of operators, are time-dependent. In this representation, the matrix elements of $X$ are related to the time derivative of the molecular orbitals $C$ so that,
\begin{equation}\label{eqn:CDeriv}
    \dot{C}_{pq}(t) = \sum_r C_{pr}(t)X_{rq}(t).
\end{equation}
We find equations for the amplitudes,
\begin{align}
	i\dot{t}_{\nu} &= \matrixel{\Phi_0(t)}{\nu\left(\bar{H} - ie^{-T}Xe^T\right)}{\Phi_0(t)}\label{eqn:OCC_dT2}\\
	-i\dot{\lambda}^{\nu} &= \matrixel{\Phi_0(t)}{(1 + \Lambda)\left[\bar{H} - ie^{-T}Xe^T,\nu^{\dagger}\right]}{\Phi_0(t)}\label{eqn:OCC_dL2},
\end{align}
and a linear equation for the orbital rotation parameters ($X$),
\begin{align}
    &\matrixel{\Phi_0(t)}{(1 + \Lambda)e^{-T}[H - i\dot{T} - iX,\alpha]e^T}{\Phi_0(t)}\nonumber \\
	&+ i\matrixel{\Phi_0(t)}{\dot{\Lambda}e^{-T}\alpha e^T}{\Phi_0(t)} + \mathrm{cc} = 0,
\end{align}
where $\alpha$ is a single excitation operator. The occupied-occupied and virtual-virtual rotations are found to be arbitrary, so that we only need to consider occupied-virtual rotations\cite{Sato2013,Sato2018CC}.

In practice, including coupled cluster singles amplitudes in the ansatz introduces a degree of redundancy that can lead to numerical problems\cite{Kvaal2012}. For this reason, we will now specialize our discussion to the case of only doubles amplitudes. This is the TD-OCCD method of Sato {\it et al}\cite{Sato2018CC}. With this simplification, the terms involving $X$ in Equations~\ref{eqn:OCC_dT2} and~\ref{eqn:OCC_dL2} vanish leading to amplitude equations that are the same as those of of TD-CCD, but with time-dependent orbitals. We may write the orbital equation as a simple linear equation,
\begin{equation}
   \sum_{bj}  A_{ia,bj}R_{bj} = b_{ia},
\end{equation}
where $i,j$ label occupied orbitals and $a,b$ virtual orbitals, and we have defined a Hermitian operator,
\begin{equation}\label{eqn:defR}
    R \equiv -iX.
\end{equation}
The $A$-matrix has a simple form in terms of the symmetrized 1RDM computed from the $T$ and $\Lambda$ amplitudes,
\begin{equation}
    A_{ia,bj} = \delta_{ab}d^j_i - d^a_b\delta_{ji},
\end{equation}
and the right-hand-side of the equation is given by,
\begin{equation}
    b_{ia} = \frac{1}{2}\left[\matrixel{L}{[H,i^\dagger a]}{R} +
    \matrixel{R}{[H,i^\dagger a]}{L}\right]
\end{equation}
where $i^\dag a$ denotes the operator for the de-excitation $a \to i$.

As mentioned previously, a consequence of including the stationary action orbital dynamics is that Ehrenfest's theorem is satisfied for all 1-electron operators. In Appendix~\ref{sec:AEhrenfest}, we show that this is true for a very general class of wavefunction Ans\"atze. 
Kadanoff and Baym introduced the notion of conserving approximations 
for many-body theories based on Green's functions. These have the property that
 the self-energies  satisfy the self-consistent Dyson equation, or equivalently, are obtained as the derivative of approximate Luttinger-Ward functionals~\cite{luttinger1960ground}. In Appendix~\ref{sec:conserving}, we briefly review such approximations. A consequence of this construction is that the resulting Green's function dynamics satisfies important single-particle conservation laws, such as for the particle density and momentum density, as well as conserve energy.
  Although the truncated coupled cluster methods are not constructed as conserving approximations, the particle density and momentum density are 1-particle properties. Thus satisfying Ehrenfest's theorem via orbital dynamics makes the zero-temperature coupled cluster dynamics ``conserving" for these properties.

%Such a self-consistent $\Sigma$ can be shown to be a functional derivative $\Sigma = \frac{\delta \Phi[G, V]}{\delta G}$, where $\Phi$ 

%{\color{red}Maybe define conserving here.}
%{\color{red}New section: do transition to expressions with indices and orbital-rotation expressions here. We should also state somewhere here that Ehrenfest's theorem is more than just about global conservation - it also improves local conservation laws, such as the continuity equation.}

\subsection{Finite-temperature coupled cluster}\label{sec:eq-cc}

%{\color{red}Include time-derivative form of Lagrangian} 
Here we review the finite temperature coupled cluster theory presented in Refs.~\onlinecite{White2018,White2020} which forms the basis of this work. This theory can be viewed as a realization of the TCC theory and is somewhat different from the thermal coupled cluster method described in Ref.~\onlinecite{Harsha2019CC}. We intend to discuss the precise relationship between various formulations of finite-temperature CC in a future work.

The theory is most compactly expressed using the imaginary time Lagrangian (Equation 5 of Ref.~\onlinecite{White2020}):
\begin{widetext}
\begin{equation}\label{eqn:FLagrangian}
	\mathcal{L} \equiv \frac{1}{\beta}\int_0^{\beta}d\tau \mathrm{E}(\tau) +
	\frac{1}{\beta}\int_0^{\beta} d\tau \lambda^{\nu}(\tau)\left[
	s_{\nu}(\tau) + \int_0^{\tau}d\tau' e^{\Delta_{\nu}(\tau' - \tau)} \mathrm{S}_{\nu}(\tau')\right].
\end{equation}
\end{widetext}
We note that this quantity is analogous to the action defined in Equation~\ref{eqn:ccaction} via its structure as a time integral. The E and S kernels resemble the ground state energy and amplitude equations respectively and are given in Appendix A of Ref.~\onlinecite{White2020}. The amplitude equations are given by
\begin{equation}\label{eqn:til_amp}
s_{\nu}(\tau) = -\int_0^{\tau}d\tau'
    e^{\Delta_{\nu}(\tau' - \tau)}
    \mathrm{S}_{\nu}(\tau'),
\end{equation}
while the equations for the $\Lambda$ amplitudes are given by
\begin{equation}
    \lambda^{\nu}(\tau) = -\mathrm{L}^{\nu}(\tau). 
\end{equation}
The L kernel is also given in Appendix A of Ref.~\onlinecite{White2020}.
Defining
\begin{equation}\label{eqn:til_lambda}
    \tilde{\lambda}^\nu(\tau)= \int_\tau^{\beta} d\tau'e^{\Delta_\nu(\tau-\tau')}\lambda^\nu(\tau'),
\end{equation}
we may rewrite the Lagrangian in the form
\begin{align}
    	\mathcal{L} &\equiv \frac{1}{\beta}\int_0^{\beta}d\tau \mathrm{E}(\tau) +
	\frac{1}{\beta}\int_0^{\beta} d\tau \lambda^{\nu}(\tau) 
	s_{\nu}(\tau) + \tilde{\lambda}^\nu(\tau)  \mathrm{S}_{\nu}(\tau).
\end{align}

%express all relevant quantities in terms of $\{s_{\nu}(\tau)\}$ and $\{\tilde{\lambda}^{\nu}(\tau)\}$ as described in Ref.~\onlinecite{White2020}. 
The boundary conditions of $s$ and $\tilde{\lambda}$ are clear from the integral equations (Equations~\ref{eqn:til_amp} and~\ref{eqn:til_lambda}),
\begin{equation}
\label{eqn:boundary}
    s(0) = 0 \quad \tilde{\lambda}(\beta) = 0.
\end{equation}
In practice, this means that the $s(\tau)$ amplitudes are propagated with the differential equation,
\begin{equation}\label{eqn:FTS}
    \frac{\partial s_{\nu}}{\partial\tau} = -\left[\Delta_{\nu} s_{\nu}(\tau)
    + \mathrm{S}_{\nu}(\tau)\right]
\end{equation}
starting from $\tau=0$ to $\tau=\beta$ along the imaginary time axis. Once the $s(\tau)$ are known in the interval $[0,\beta]$, the $\tilde{\lambda}$ amplitudes are computed according to,
\begin{equation}\label{eqn:FTL}
    \frac{\partial \tilde{\lambda}^{\nu}}{\partial \tau} = 
    \left[ \Delta_{\nu} \tilde{\lambda}^{\nu}(\tau) + 
    \mathrm{L}^{\nu}(\tau)\right]
\end{equation}
from $\tau=\beta$ back to $\tau=0$. Using these differential relations, it is insightful to re-express the Lagrangian as
\begin{align}
	\mathcal{L} &\equiv \frac{1}{\beta}\int_0^{\beta}d\tau \mathrm{E}(\tau) +
	\frac{1}{\beta}\int_0^{\beta} d\tau [(-\partial_\tau + \Delta_\nu) \tilde{\lambda}^{\nu}(\tau)] 
	s_{\nu}(\tau) \notag \\
	&+ \tilde{\lambda}^\nu(\tau)  \mathrm{S}_{\nu}(\tau) \notag\\
	&= \frac{1}{\beta}\int_0^{\beta}d\tau \mathrm{E}(\tau) +
	\frac{1}{\beta}\int_0^{\beta} d\tau \tilde{\lambda}^\nu(\tau) \left[(\partial_\tau + \Delta_\nu) 	s_{\nu}(\tau) + \mathrm{S}_{\nu}(\tau) \right]
\end{align}
where we have integrated by parts and used the boundary conditions in Equation~\ref{eqn:boundary}, and the latter
expression is analogous to the form of action in Equation~\ref{eqn:ccaction}, with $\tilde{\lambda}^\nu$ playing the role of the $\Lambda$ amplitudes in that equation.

As with the zero-temperature theory, finite-temperature coupled cluster is not ``conserving" in the sense of Kadanoff and Baym\cite{Baym1961,Baym1962}. The equilibrium theory nonetheless provides a framework to compute properties as derivatives
of the grand potential. Details relating to this response formulation of properties are discussed in Ref.~\onlinecite{White2020}.

\subsection{Coupled cluster dynamics at finite temperature}\label{sec:ft_cc_dynamics}

\begin{figure}
    \centering
    \includegraphics[scale=1.0]{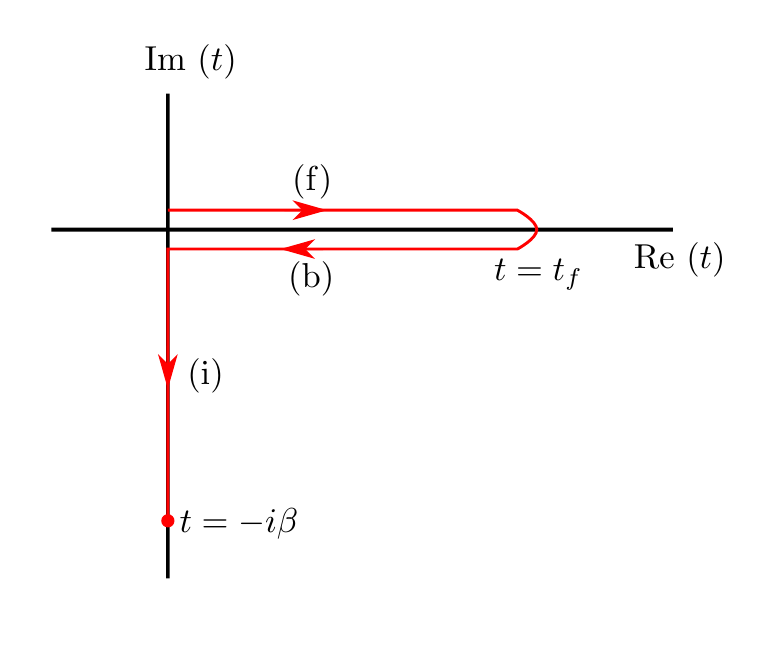}
    \caption{The Keldysh contour including the imaginary branch. The propagation proceeds forwards in real time, backwards in real time, and in imaginary time respectively. These three branches are labelled as (f), (b), and (i). Contours like this which include the imaginary time branch first appeared in the work of Konstantinov and Perel'\cite{Konstantinov1960}.}
    \label{fig:contour_kel}
\end{figure}

%{\color{red}Keldysh needs to be recapped}
FT-CC theory can be extended to treat out-of-equilibrium systems using the Keldysh formalism as discussed in Ref.~\onlinecite{White2019}. The key idea is to analytically continue the imaginary time formalism of Section~\ref{sec:eq-cc} onto the real axis using a contour like the one shown in Figure~\ref{fig:contour_kel}. Computation of the response density matrices along this contour provide observables of a thermal system driven out of equilibrium as discussed in Appendix A of Ref.~\onlinecite{White2019}. Extending the equilibrium Lagrangian onto the Keldysh contour yields
\begin{widetext}
\begin{equation}\label{eqn:Lcc}
	\mathcal{L} \equiv \frac{i}{\beta}\int_C dt \mathrm{E}(t) +
	\frac{i}{\beta}\int_C dt \lambda^{\nu}(t)\left[
	s_{\nu}(t) + i\int_{C(0)}^{C(t)}dt' e^{i\Delta_{\nu}(t' - t)} \mathrm{S}_{\nu}(t')\right].
\end{equation}
\end{widetext}
where the integral over $t'$ is understood to proceed along the contour. This Lagrangian leads to the same differential equations, now expressed along the real axis,
\begin{equation}\label{s}
\begin{split}
    \dot{s}_\nu(t)=(-i)\left[\Delta_\nu s_\nu(t) + \mathrm{S}_\nu(t)\right]
\end{split}
\end{equation}
\begin{equation}\label{l}
\begin{split}
    \dot{\Tilde{\lambda}}(t)=i\left[\Delta_\nu\Tilde{\lambda}^\nu(t) + \mathrm{L}^\nu(t)\right].
\end{split}
\end{equation}
These equations are the analytic continuation of Equations~\ref{eqn:FTS} and~\ref{eqn:FTL}. 
When $\nu$ is restricted to singles and doubles, these equations define the Keldysh-CCSD method.
Using the differential relations, we can similarly rewrite the Lagrangian as
\begin{align}
\label{eqn:ccftdaction}
    \mathcal{L} \equiv \frac{i}{\beta}\int_C dt \mathrm{E}(t) +
	\frac{i}{\beta}\int_C dt \tilde{\lambda}^{\nu}(t)\left[(-i\partial_t + \Delta_\nu)s_\nu(t) + \mathrm{S}_\nu(t)\right].
\end{align}

\begin{figure}
    \centering
    \includegraphics[scale=1.0]{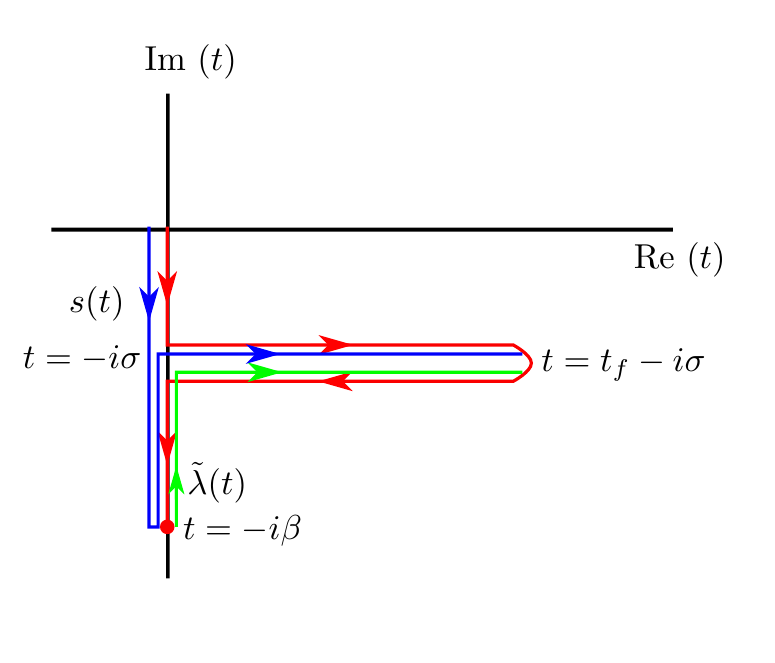}
    \caption{The contour used in this work (red) with propagation directions for $s$ and $\tilde{\lambda}$ shown in blue and green respectively. Note that some branches of the contour are drawn slightly away from the actual contour so that all branches can be shown clearly.}
    \label{fig:contour_prop}
\end{figure}

In practical calculations, there is a degree of freedom in choosing the position of the real branch of the contour relative to the imaginary-time integration limits. This corresponds to the choice of $\sigma$ in Figure~\ref{fig:contour_prop}. Furthermore, once $s$ and $\tilde{\lambda}$ are known along the whole of the imaginary branch of the contour, they can be propagated simultaneously in real time, starting from $t = -i\sigma$. In effect, $s$ is propagated forward along the forward branch of the contour and $\tilde{\lambda}$ is propagated backwards along the reverse branch of the contour so that at each time the density matrices for the non-equilibrium system can be constructed.

%\blue{In practical calculation using the differential versions \eqref{s},\eqref{l}, the $s(t)$ and $\tilde{\lambda}(t)$ amplitudes are simultaneously propagated from $t=0$ to $t=t_f$, unlike the calculation of equilibrium amplitudes described in Section~\ref{sec:eq-cc}, as shown in Figure~\ref{fig:contour_prop}. Furthermore, the time variable in Equation \eqref{s},\eqref{l} are both the same time on the forward branch (f) in the conventional Keldysh contour notation in Figure~\ref{fig:contour_kel}. (I find it hard to explain, but the idea is that the $s$ directions in Fig.~\ref{fig:contour_prop} is always the same direction as the red arrows in Fig.~\ref{fig:contour_kel}, whereas $\tilde{\lambda}$ direction is always reverse of the red arrows.) }  

%\begin{figure}[h!]

%\begin{subfigure}{0.5\textwidth}
%	\centering
%	\includegraphics[]{contour_kel.jpeg}
%	\caption{The conventional Keldysh contour.}
%	\label{fig:contour_kel}
%\end{subfigure}

%\begin{subfigure}{0.5\textwidth}
%	\centering
%	\includegraphics[]{contour_prop.png}
%	\caption{Practical propagation directions. }
%	\label{fig:contour_prop}
%\end{subfigure}

%\caption{ }
%\label{fig:contours}
%\end{figure}

%{\color{red}Define conserving approximation here.}
The Keldysh-CC method has
two deficiencies which appear when  propagating for longer times. (i) As is the case for zero-temperature CC dynamics, Keldysh-CC truncated to singles and doubles does not in general satisfy Ehrenfest's theorem. (ii) The time-dependent properties are not stationary under propagation with the same time-independent Hamiltonian that generates the equilibrium density matrix. (However, the energy remains stationary under propagation by a time-independent Hamiltonian, for the same reasons that it is conserved in zero-temperature coupled cluster dynamics). 
 (i) was discussed already in Ref.~\onlinecite{White2019} which, in particular, showed analytically and numerically that  approximate Keldysh-CC dynamics violates global particle number conservation. 
As suggested by previous arguments in Section~\ref{sec:zt-orb-dyn}, (i) can be addressed by introducing stationary orbital dynamics.
(ii) is a different problem, and we will discuss the ``stationarity of equilibrium approximations" in Section~\ref{sec:stationary}. 

\subsection{Orbital-dependent FT-CC dynamics}\label{sec:orb_cc_dynamics}

We now extend the formulation of Section~\ref{sec:ft_cc_dynamics} to a time-dependent orbital basis with the goal of satisfying Ehrenfest's theorem for all 1-particle observables at finite temperature. As in the zero-temperature case, we re-express the time-dependence of the orbitals via the time-dependence of the matrix elements of operators in the orbital basis.
%can be equivalently formulated as time-dependence of the integrals. 
For example, the representation of the 1-particle part of the Hamiltonian undergoes dynamics as
\begin{equation}\label{orb-update}
    h_{pq}(t) = \sum_{rs} h_{rs}C_{r p}^*(t)C_{s q}(t),
\end{equation}
where here $r$ and $s$ refer to the starting orbitals.
%{\color{red}Move comment above.}
In analogy with the  zero-temperature theory, we define an $X$ operator (see Equation~\ref{eqn:CDeriv}) for non-redundant pairs, %\cite{Sato2018CC,Sato2013,Kretchmer2018}
$X_{ai}(t)=-X_{ia}(t)^*$. In the finite-temperature case, recall that $i,j,\ldots$ and $a,b,\ldots$ are used to refer to orbitals with hole and particle character respectively, but these orbitals will span the full space in general. (In principle, we can introduce the matrix elements in the pure-hole and pure-particle sector, $X_{ij}(t)$, $X_{ab}(t)$, but as shown in Appendix~\ref{sec:AEhrenfest}, by the choice of Lagrangian below, such elements vanish). As in the zero-temperature case, the introduction of orbital rotations eliminates the need for singles amplitudes. We will refer to the resulting doubles theory as Keldysh-OCCD.

To derive the equations, we define a symmetrized Lagrangian,
\begin{equation}\label{full_Q}
    \mathcal{Q}
     \equiv \frac{1}{2}(\mathcal{L}+\mathcal{L}^*) + 
    \Omega^{(0)} + \frac{i}{\beta}\int_C dt \mathrm{E}^{(1)}(t).
\end{equation}
The Lagrangian is symmetrized to ensure that the $X$ matrix is anti-Hermitian, and the second and third terms correspond to the thermal Hartree-Fock contribution.
%\begin{equation}
%    \mathrm{E}^{(0)}(t) + \mathrm{E}^{(1)}(t) = h_{ii}(t) + \frac{1}{2}\langle ij||ij\rangle(t).
%\end{equation}
$\mathcal{L}$ has a similar form to Equation \ref{eqn:ccftdaction} (or equivalently Equation \ref{eqn:Lcc}) but we choose to remove the term containing $\Delta_\nu$ (further discussion below), 
\begin{align}\label{theory1_L}
        \mathcal{L} \equiv \frac{i}{\beta}\int_C dt \mathrm{E}(t) +
	\frac{i}{\beta}\int_C dt \tilde{\lambda}^{\nu}(t)\left[(-i\partial_t)s_\nu(t) + \mathrm{S}_\nu(t)\right].
\end{align}
In addition, 
 the kernels $\mathrm{E}[s(t)]$, $\mathrm{S}_\nu[s(t)]$, $\mathrm{L}^\nu[s(t),\tilde{\lambda}(t)]$ are modified as follows: 
(i)  all Hamiltonian tensors are now time-dependent; (ii) the time-derivative of the orbitals results in a modification of the 1-electron integrals\cite{Sato2018CC,Sato2013}
\begin{equation}\label{eqn:formal_R}
h_{ai} \rightarrow h_{ai}(t) - iX_{ai}(t)
\end{equation}
and similarly for $h_{ia}$. As in Ref.~\onlinecite{White2020}, our notation includes factors of the square root of the occupation numbers in the definition of the tensors. For example,
\begin{equation}\label{eqn:integrals}
    h_{ai} \equiv \sqrt{n_i\bar{n}_a}\matrixel{a}{h}{i}
\end{equation}
 (iii) the Fock matrix becomes time-dependent and is computed from the time-dependent Hamiltonian tensors, and unlike in Keldysh-CC, the diagonal is not subtracted (further discussed below)
%\begin{equation}\label{eqn:subtract_zero_oo}
%    f_{ij} = h_{ij}(t) + \langle ik||jk\rangle(t) - \sqrt{n_in_j}\delta_{ij}\varepsilon_i
%\end{equation}
%\begin{equation}\label{eqn:subtract_zero_vv}
%    f_{ab} = h_{ab}(t)+\langle ak||bk\rangle(t)-\sqrt{\Bar{n}_a\Bar{n}_b}\delta_{ab}\varepsilon_a
%\end{equation}\
\begin{align}\label{eqn:subtract_zero_oo}
    f_{ij} &= h_{ij} + \langle ik||jk\rangle 
    - \sqrt{n_in_j}\delta_{ij}\varepsilon_i \ \ &\text{Keldysh-CC} \notag\\
    \to
        f_{ij}(t) &= h_{ij}(t) + \langle ik||jk\rangle(t) \ \ &\text{Keldysh-OCC}
\end{align}
\begin{align}\label{eqn:subtract_zero_vv}
    f_{ab} &= h_{ab}+\langle ak||bk\rangle
    -\sqrt{\Bar{n}_a\Bar{n}_b}\delta_{ab}\varepsilon_a \ \  &\text{Keldysh-CC} \notag\\
    \to
    f_{ab}(t) &= h_{ab}(t)+\langle ak||bk\rangle(t) \ \ &\text{Keldysh-OCC}
\end{align}
where in the Keldysh-CC formulation,  $\varepsilon_a$, $\varepsilon_i$ are orbital energies, and  $n_i,\Bar{n}_a=1-n_a$ are orbital occupancies. 

In the Keldysh-CC equations, the $\Delta_\nu$ term in the Lagrangian and the corresponding subtraction of occupancy weighted eigenvalues from the Fock operator both originate from the choice of zeroth order Hamiltonian, and together ensure that the diagrams of Keldysh-CC correspond to a well-defined time-dependent perturbation theory. The mathematical effect of these terms is to introduce eigenvalue time-dependent phases on the indices of the amplitudes during the propagation, as seen from the amplitude equations \ref{s}, \ref{l}. In the orbital-optimized Keldysh-CC, such time-dependent phases are fully determined by stationarity of the Lagrangian with respect to the rotation elements $X_{ij}, X_{ab}$. Thus we can remove both terms involving the zeroth order eigenvalues discussed above, while obtaining the same result at stationarity. As shown explicitly in Appendix~\ref{sec:AEhrenfest} this choice leads to $X_{ij}, X_{ab}=0$.

%We denote the kernels with the above modification as
%\begin{equation}
%    \mathrm{E}[H_1(t)-iX(t);s(t)] = \mathrm{E}[H'_1(t);s(t)]
%\end{equation} 
%and similarly for $\mathrm{S}_\nu(t),\mathrm{L}^\nu(t)$. Furthermore, to simplify notation, we will drop the amplitude argument in the square bracket, so that the modified kernels are simply represented as $\mathrm{E}[H'_1(t)]$, and so forth. The kernels of FT-CC with time-independent orbitals are represented as $\mathrm{E}[H_1(0)]$ using this notation. 

To obtain the amplitude equations, we set the variations of $\mathcal{Q}$ with respect to the amplitudes to zero, which is equivalent to setting the amplitude variations of $\mathcal{L}$ to zero. Since all terms containing singles amplitudes vanish, $f_{ia}$ and $f_{ai}$ do not enter the kernels directly. Therefore, as in the analogous zero-temperature theory, the appearance of $X_{ai}$ in $h_{ai}(t)$ does not affect the Keldysh-OCCD amplitude equations. This leads to Equations~(\ref{s}) and (\ref{l}) but with the S and L kernels containing time-dependent matrix elements, and the $\Delta_\nu$ contribution missing
    \begin{equation}
\begin{split}
    \dot{s}_\nu(t)=-i \mathrm{S}_\nu(t)
\end{split}
\end{equation}
\begin{equation}
\begin{split}
    \dot{\Tilde{\lambda}}(t)=i \mathrm{L}^\nu(t).
\end{split}
\end{equation}
Precise equations for the kernels are given in Appendix~\ref{app:equations}.

%(ii) the $H^0$ subtracted from the Fock matrices in the kernels is a constant, and thus has a vanishing time derivative and variation; (iii) 
%The Lagrangian is symmetrized to ensure that the $X$ matrix is anti-Hermitian. Finally,
%\begin{equation}
 %   \mathrm{E}^{(0)}(t) + \mathrm{E}^{(1)}(t) = h_{ii}(t) + \frac{1}{2}\langle ij||ij\rangle(t)
%\end{equation}
%corresponds to the time-dependent thermal Hartree-Fock energy.

To obtain an equation for $X$, we vary $\mathcal{Q}$ with respect to the orbital parameters and set the resulting expression to zero. As in the zero-temperature case, this yields a linear equation of the form
\begin{equation}\label{eqn:ft_orb}
    \sum_{bj} A_{ia,bj}R_{bj} = b_{ia},
\end{equation}
where we again use the Hermitian $R$ defined in Equation~\ref{eqn:defR}. The elements of these tensors are found to be
\begin{equation}
    A_{ia,bj} = \delta_{ab}d_{ij} - d_{ba}\delta_{ji}
\end{equation}
and
\begin{equation}
    b_{ia} = \mathcal{F}_{ai} - \mathcal{F}_{ia}^{\ast}.
\end{equation}
Here we have used $d_{pq}$ to represent elements of the symmetrized reduced density matrices and
\begin{equation}\label{eqn:orbF}
\begin{split}
    \mathcal{F}_{pq}
    \equiv d_{pr}h_{rq}+\frac{1}{2} \sum_{su} d^{ps}_{uv} \langle uv||qs\rangle.
\end{split}
\end{equation}
where $d^{ps}_{uv}$ is an element of the 2-particle reduced density matrix (2RDM). As in the zero temperature case, the inclusion of such orbital dynamics leads to the satisfaction of Ehrenfest's theorem for all 1-particle properties, as shown in Appendix~\ref{sec:AEhrenfest}.

\subsection{Stationarity of equilibrium approximations}\label{sec:stationary}
\begin{figure}
    \centering
    \includegraphics[scale=0.6]{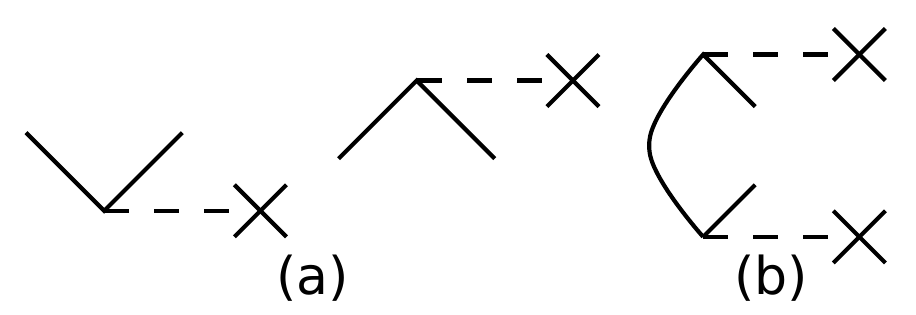}
    \caption{Diagrammatic contributions to the PT2 1-RDM for a 1-particle problem (a), including the contribution from the the response of the reference (b).}
    \label{fig:PT2}
\end{figure}

In general, there is no guarantee that an approximate equilibrium density matrix will be stationary when propagated in real-time with the equilibrium Hamiltonian. This property holds for  conserving approximations because the approximate Luttinger-Ward functional is expressed in terms of Feynman diagrams, where each diagram implicitly includes all time-orderings  of the interactions. For a more general class of theories, this suggests that a necessary condition for stationarity is that all time-ordering counterparts of a particular time-ordered diagrammatic contribution to the density matrix should be included in the theory. This is clearly violated in approximate coupled cluster theory. Perturbation theory at finite order includes all time-orderings at a particular order which means that the density defined strictly as the derivative of the action with respect to an external potential has this property (see  Appendix~\ref{app:stationary} for a more detailed demonstration). 
For example, consider perturbation theory at 2nd order (PT2) with a 1-particle perturbation. In Figure~\ref{fig:PT2} (a) we show the diagrammatic contributions to the PT2 one-particle density matrix obtained by differentiating the energy expression with respect to the applied potential. 
The density matrix defined in this way includes both time orderings of the relevant diagram
and has the stationary property. 
However, one often includes a contribution from the response of the reference via terms like those in Figure~\ref{fig:PT2} (b). In this case, not all time-orderings of the same diagrammatic contribution are included and the density is no longer stationary when propagated in real time. 

In the case of perturbation theory, this property can be restored by including some higher order terms. For coupled cluster theory with limited excitations (such as truncation to singles and doubles) this is not generally possible, and, as we have stated previously, none of the methods discussed in this work have this stationarity property. For short or moderate propagation times this can be corrected by subtracting the anomalous dynamics. To be precise, given a Hamiltonian of the form
\begin{equation}
    H(t) = H_0 + V(t) \quad [V(0) = 0],
\end{equation}
we can define a corrected density matrix
\begin{equation}
    \rho_c(t) = \rho(t) + \rho(0) - \rho_0(t)
\end{equation}
where $\rho(t)$ is the density propagated with the full $H(t)$ and $\rho_0(t)$ is the density propagated with the time-independent Hamiltonian $H_0$.

\section{Implementation}\label{sec:implementation}

In the following Section~\ref{sec:result}, the Keldysh-CCSD results are obtained using the implementation as described in Ref.~\onlinecite{White2019}. The Keldysh-OCCD results are obtained as follows: (i) the equilibrium amplitudes $s(\tau),\tilde{\lambda}(\tau)$ are first computed as described in Section~\ref{sec:eq-cc}. In particular, the differential form of the $s$-amplitude equation (Equation \ref{eqn:FTS}) is first propagated from $\tau=0$ to $\tau=\beta$ along the imaginary time contour. The $s(\tau)$ amplitudes are then used in Equation \ref{eqn:FTL} to propagate $\tilde{\lambda}(t)$ from $\tau=\beta$ back to $\tau=0$.   These equilibrium amplitudes are computed within the coupled cluster doubles approximation, using fixed orbitals, and without any truncation of the occupied or virtual space. The 4th order Runge-Kutta scheme is used to propagate the differential equations. (ii) The equilibrium amplitudes at $\tau=\beta/2$ are taken as the initial $t=0$ amplitudes for the subsequent dynamics. This corresponds to a choice of contour where the real part extends from $-i\beta/2$ in Figure~\ref{fig:contour_prop}. (iii) The dynamical $s(t)$ and $\tilde{\lambda}(t)$ amplitudes are computed as described in Section~\ref{sec:ft_cc_dynamics}. In particular, Equation \ref{s}, \ref{l} are simultaneously propagated from $t=0$ to $t=t_f$ on the real contour where the kernels $S$, $L$ are modified to include orbital dynamics as described in Section~\ref{sec:orb_cc_dynamics}. 
%{\color{red}\sout{In the differential equation, it looks like $\lambda(t)$ depends on $s(t)$, so you need the value on the same branch of the contour, if $t$ is contour time. If $t$ is not contour time, then that needs to be clarified.}} {\color{red}\sout{The next observation should be made I think in the theory section, see previous comment.}} 
%\sout{Note that unlike the equilibrium amplitudes where $s(\tau)$ is computed first on the entire imaginary branch from $\tau=0$ to $\tau=\beta$ and then $\tilde{\lambda}(\tau)$ propagated back from $\beta$ to $0$ using the $s(\tau)$ amplitudes obtained at each $\tau$, $s(t)$ and $\tilde{\lambda}(t)$ on the real branch are updated simultaneously both from $t=0$ to some final time $T$.} 
(iv) At each update of $s(t)$ and $\tilde{\lambda}(t)$, the orbital Equation \ref{eqn:ft_orb} is solved to obtain $R$, which is used to update the integrals according to e.g. equation \ref{orb-update}. We use the 4th order Runge-Kutta (RK4) scheme to update both the amplitudes and the orbitals in a coupled manner as described below. 
The amplitude Equations \ref{s}, \ref{l} and orbital coefficient Equation \ref{eqn:CDeriv} are of the form 
\begin{equation}
\begin{split}
\frac{dy}{dt}&=f(t,y(t),h(C(t))) \\
\frac{dC}{dt}&=C X(t,d(y(t)),h(C(t))) 
\end{split}
\end{equation}
%\begin{align}
%\end{align}
where $y \in \{ s(t), \tilde{\lambda}(t)\}$, $h(C(t))$ denotes the Hamiltonian integral matrix elements in the time-dependent orbital basis $C(t)$, and $d(y(t))$ denotes the reduced density matrices computed from the amplitudes at time $t$. 
In RK4, the time-step from $t \to t+\delta t$ is assembled from intermediate amplitudes $y_i$ and orbitals $C_i$ as well as their respective finite difference $\delta y_i$ and $X_i$ at intermediate times $t_i$, for $i=1\ldots 4$. The intermediate quantities are computed as
\begin{equation}
\begin{split}
\delta y_i & = f(t_i,y_i,h(C_i))\\
X_i & = X(t_i, d(y_i), h(C_i))
\end{split}
\end{equation}
where
\begin{equation}
\begin{split}
t_i & = t+a_i\delta t\\
y_i & = y(t)+a_i\delta t\delta y_{i-1}\\
C_i & = C(t)e^{a_i\delta tX_{i-1}}
\end{split}
\end{equation}
%\begin{align}
% \delta y_i &=f(t+a_i\delta t,y+a_i\delta t\delta y_{i-1}, h_{i-1}) \nonumber\\
% h_i &= h_0 (C_i) \nonumber\\
%C_i &= C_0 e^{-i a_i \delta t R_{i-1} } \\
%R_{i-1} & = R()
%\end{align}
where $a_i$ are standard 4th order Runge-Kutta weights $\{ 0, \frac{1}{2}, \frac{1}{2}, 1 \}$. 
%, and the second line indicates the transformation of the one- and two-particle integral tensors of $h_0$. , with 
Note that all quantities for $i=1$ correspond to their values at time $t$.

%The intermediate 1- and 2-RDMs are then computed from the intermediate  amplitudes $y+a_i\delta t\delta y_i$ {\color{red}is this consistent with the above equation}, which are then used to compute the intermediate orbital rotation parameters $R_i$, which then determine the Hamiltonian tensors for the next intermediate time $i+1$. 
After all intermediate quantities are obtained, the amplitudes and orbitals are updated as 
\begin{align}
y(t+\delta t)&=y(t)+\frac{1}{6}\delta t(\delta y_1+2\delta y_2+2\delta y_3+\delta y_4)\notag\\
X(t+\delta t)&=\frac{1}{6}(X_1+2X_2+2X_3+X_4)\notag\\
C(t+ \delta t) &= C(t) e^{\delta t X(t+\delta t)}
\end{align}

The present Keldysh-OCCD theory has the same scaling as the previous Keldysh-CCSD theory in terms of computational cost and memory. The amplitude update cost scales as $N^6$ for each real or imaginary time step, where $N$ is the size of 1-particle basis. Furthermore for each real time step, Keldysh-OCCD additionally involves solving a linear set of equations for $N^2$ variables in
%inverting a matrix of dimension $(N^2,N^2)$ when solving 
the orbital equation \ref{eqn:ft_orb}, and an integral update which scales as $N^5$. 

\section{Results}\label{sec:result}

\subsection{Numerical demonstration of Ehrenfest's theorem}

\begin{figure}[h]
    \centering
    \includegraphics[]{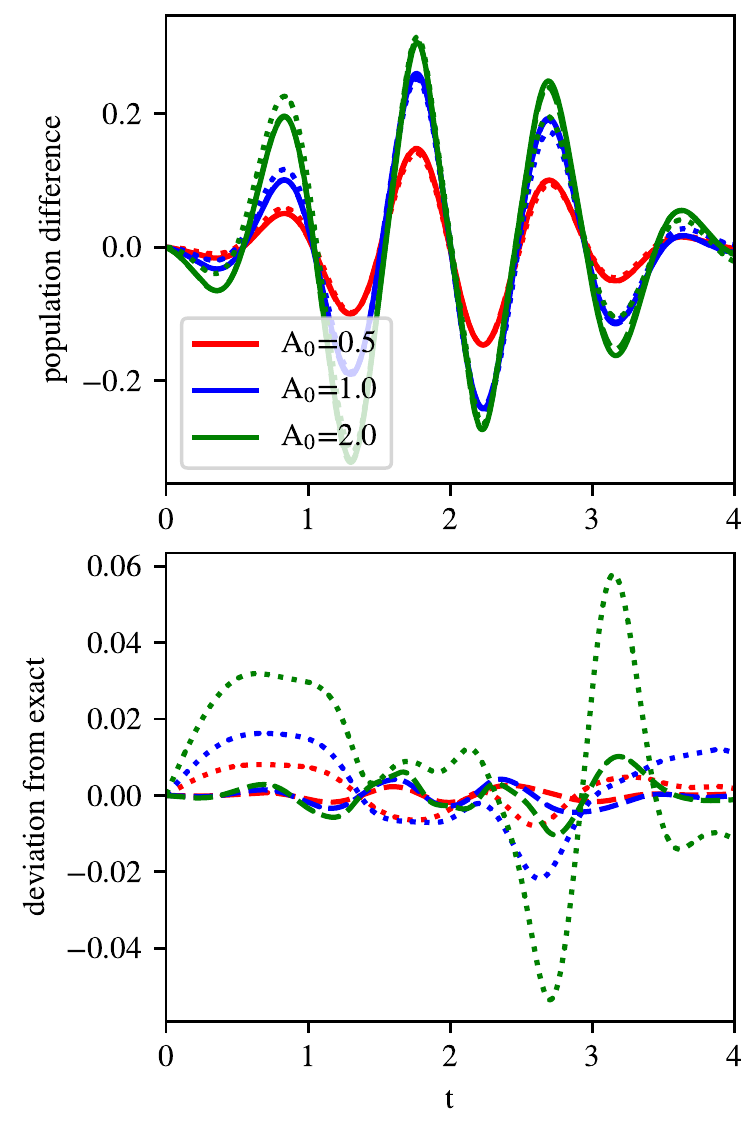}
    \caption{Population difference $n_L - n_R$ as a function of time for the 2-site Hubbard model at half-filling. The solid line is the exact result, the dashed line is the Keldysh-OCCD result and the dotted line is the real part of the Keldysh-CCSD result. In the lower panel, we show the difference of Keldysh-OCCD and Keldysh-CCSD results from the exact result.}
    \label{fig:hubbard_hf}
\end{figure}

We first demonstrate the satisfaction of Ehrenfest's theorem for the Keldysh-OCCD method for the 2-site time-dependent Peierls-Hubbard model. This was previously studied with Keldysh-CCSD in  Ref.~\onlinecite{White2019}. The time-dependent Hamiltonian is given by 
\begin{equation}
    H(t)=-t_H\sum_{i\sigma}\left[e^{iA(t)}a^\dagger_{i\sigma}a_{(i+1)\sigma}+h.c.\right]+U\sum_in_{i\uparrow}n_{i\downarrow}
\end{equation}
where the first term is the Peierls driving term, which mimics the effect of the coupling of an underlying nuclear lattice to an external laser pulse. Here, the driving takes the form
\begin{equation}
    A(t)=A_0e^{-(t-t_0)^2/2\sigma^2}\cos[\omega(t-t_0)].
\end{equation}
We use $U=1.0$, for which a chemical potential of $\mu=0.5$ gives half filling at equilibrium, a temperature $T=1.0$, and pulse parameters of $\sigma=0.8$, $t_0=2$, $\omega=6.8$. We show the Keldysh-CCSD and Keldysh-OCCD results for the population difference between the 2 sites ($L, R$) defined as $n_L-n_R$, where $n_L=n_{L,\uparrow}+n_{L,\downarrow}$, along with the exact result computed by propagating the density matrix in the full Liouville space.
As shown in Figure \ref{fig:hubbard_hf}, the deviation from the exact result increases for both Keldysh-CCSD and Keldysh-OCCD with increasing pulse amplitude $A_0$. However, the orbital dynamics in Keldysh-OCCD ensures that it is much closer to the exact result for all pulse parameters.

\begin{figure}[h]
    \centering
    \includegraphics[]{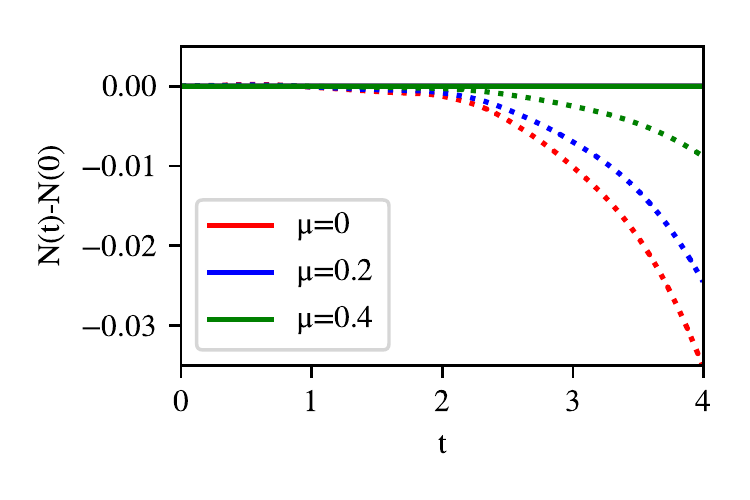}
    \caption{Change of total particle number $N$ in the 2-site Peierls-Hubbard model with different chemical potentials $\mu$. The solid line is the Keldysh-OCCD result and the dotted line is the real part of the Keldysh-CCSD result. Keldysh-OCCD satisfies the global particle number conservation law. }
    \label{fig:hubbard_mu}
\end{figure}

As discussed in detail in Ref.~\onlinecite{White2019}, Keldysh-CCSD does not in general conserve global symmetries, such as the total particle number, whereas such 1-particle quantities are conserved both locally and globally by Keldysh-OCCD. In Figure~\ref{fig:hubbard_mu}, we show the change of the total particle number of Keldysh-CCSD and Keldysh-OCCD for the 2-site Hubbard model for $\mu\ne U/2$ where the particle-hole symmetry is no longer present. As expected, we see that as $\mu$ is decreased from half-filling, the Keldysh-CCSD total particle number begins to deviate from the equilibrium value at longer times, whereas the total particle number remains conserved for Keldysh-OCCD for each $\mu$. 

\begin{figure}[h]
    \centering
    \includegraphics[]{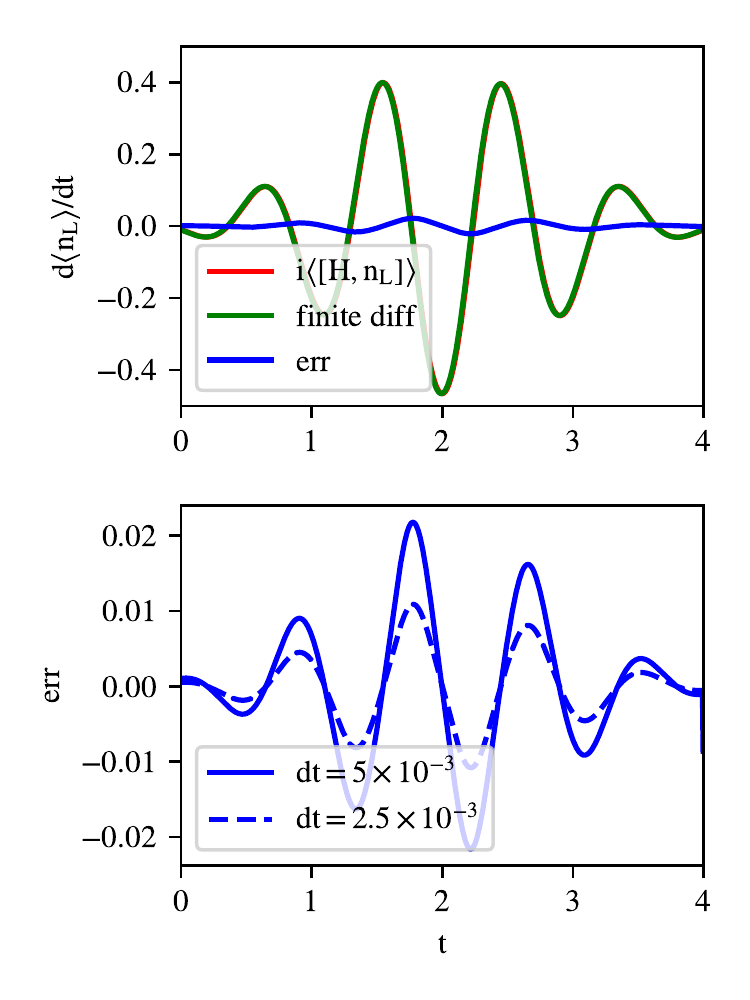}
    \caption{Time-derivative of the Keldysh-OCCD left-site population $d \langle n_L\rangle/dt$ in the 2-site Peierls-Hubbard model for $\mu=0$. The red curve in the upper panel is the population derivative computed from the flux $i\langle [H, n_L]\rangle$ and the green curve is the numerical time-derivative using a finite difference time-step of $dt=5 \times 10^{-3}$. The blue curve is the difference between the flux expression and the finite difference time-derivative; this should be zero if Ehrenfest's theorem is satisfied. The lower panel plots the same but for two time-steps of $dt=5\times 10^{-3}$ and $dt=2.5\times 10^{-3}$. These results show that Keldysh-OCCD satisfies Ehrenfest's theorem for the local particle number, and thus the local particle number conservation law.}
    \label{fig:hubbard_local}
\end{figure}

In Figure~\ref{fig:hubbard_local}, we demonstrate the stronger condition of conservation of local particle number for the population of the left site. From Ehrenfest's theorem, 
\begin{equation}\label{eqn:local}
\frac{d}{dt}\langle n_L\rangle=i\langle[H,n_L]\rangle
\end{equation}
where the expression on the right hand side is the contribution from the right-site flux. The red curve in the upper panel shows the local population change computed with the flux. The green curve gives the time-derivative of $n_L$, computed by the finite-difference expression $(\langle
n_L\rangle(t+\delta t)-\langle n_L\rangle(t))/\delta t$, with time step $5\times 10^{-3}$. The two curves are on top of each other, thus we also plot the difference between the r.h.s. and the l.h.s. quantities. This stays close to $0$ at all times, demonstrating the local conservation law. In the lower panel, we also plot the difference for a smaller time-step of $2.5\times 10^{-3}$ where the quantity is reduced by a factor of 2. This illustrates that Ehrenfest's theorem will be fully satisfied for an infinitesimal time step.

%In the lower panel, the difference between the two sides of Equation~\ref{eqn:local} is plotted for time step 5e-5 as the solid curve and 2.5e-5 as the dashed curve. We see that reducing the time step by half also reduce the error by half, thus numerically demonstrating the conservation. 
%{\color{red}\sout{Overall this is not a full demonstration of Ehrenfest's theorem, as it is just showing global conservation. Can we instead show the local conservation, e.g. rate of change of density on a site and $[H, n_0]$ and show that they are equal?}}

\subsection{Comparison with Keldysh-CCSD: warm-dense Si}

We next compare the performance of Keldysh-CCSD and Keldysh-OCCD for the field-driven Si system in Ref.~\onlinecite{White2019}. This treats a single primitive cell of Si in a minimal basis (SZV\cite{VandeVondele2007} with GTH-Pade pseudopotentials\cite{Goedecker1996,Hartwigsen1998}) and with the ions frozen at the experimental lattice constant (3.567\r{A}). The matrix elements were obtained from the PySCF software package using plane-wave density fitting\cite{PYSCF,sun2020recent,sun2017gaussian}. We use the dipole approximation in the velocity gauge where the coupling to an external field is of the form
\begin{equation}
    \frac{1}{mc}\Vec{p}\cdot\Vec{A}(t)
\end{equation}
where $\Vec{p}$ is the momentum operator, and we choose a pulse shape described by 
\begin{equation}\label{pulse}
    \frac{1}{c}\Vec{A}(t)=A_0\Vec{z}e^{-(t-t_0)^2/2\sigma^2}\cos[\omega(t-t_0)].
\end{equation}
We choose parameters of $\sigma=2.0$, $t_0=15.0$, $\omega=0.97529$, $A_0=0.6752$ at a temperature $T=0.2$. This corresponds to a maximum laser intensity of $2.36\times10^{16}$ W/cm${}^2$.

\begin{figure}[h]
    \centering
    \includegraphics[]{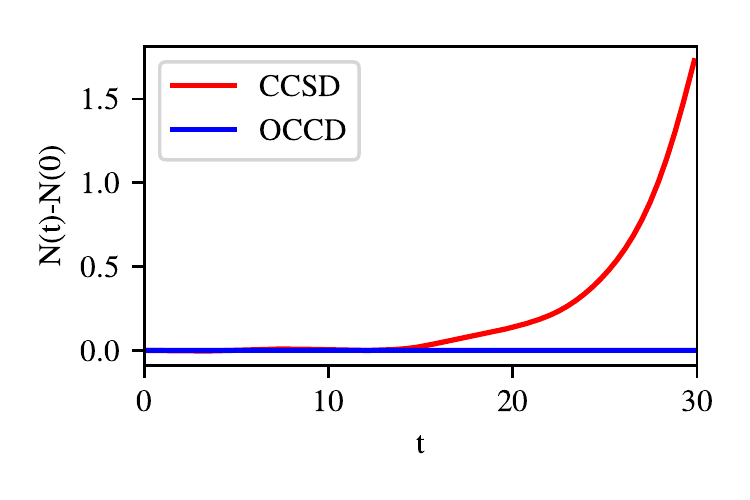}
    \caption{Change of the number of electrons per unit cell for the Si system as a function of time. }
    \label{fig:Si_total}
\end{figure}

\begin{figure}[h]
    \centering
    \includegraphics[]{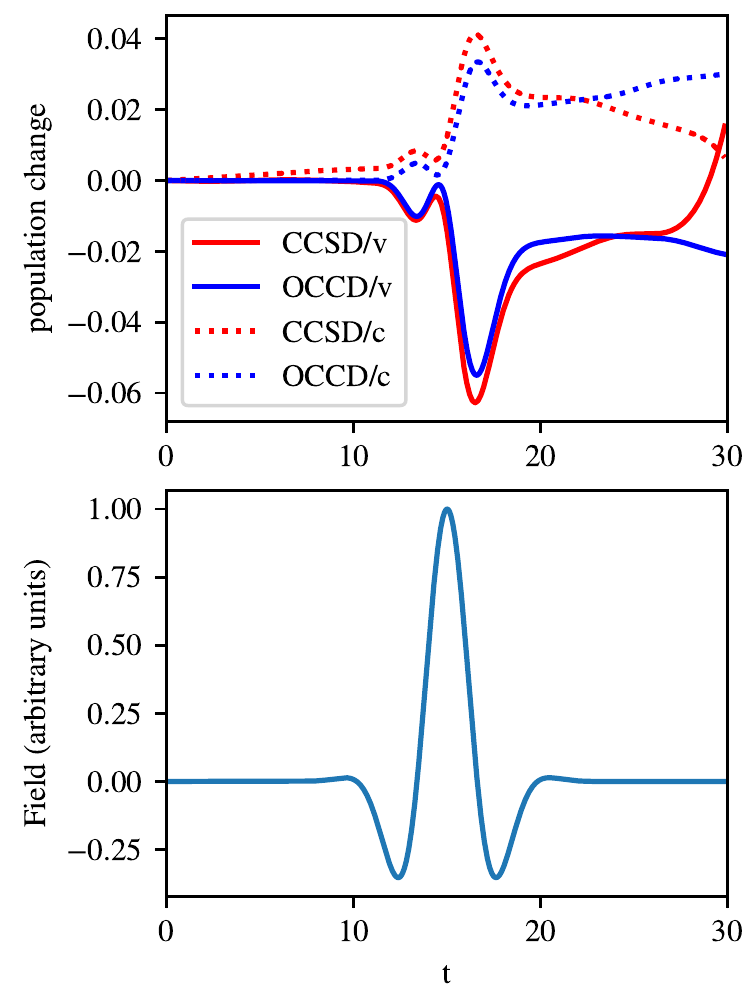}
    \caption{Difference in population of the valence (solid line) and conduction band (dotted line) for Keldysh-CCSD and Keldysh-OCCD as a function of time. The shape of the electric field as a function of time is plotted in the lower panel. Additional discussion in the text.}
    \label{fig:Si_vc}
\end{figure}

Figure \ref{fig:Si_total} again demonstrates that, as expected, Keldysh-OCCD conserves total particle number in contrast to Keldysh-CCSD. 
%This fixes one of the essential problems with the Keldysh theory presented in Ref.~\onlinecite{White2019}. 
In Figure \ref{fig:Si_vc} we show the change in the population of the valence and conduction bands induced by the laser. While the Keldysh-CCSD results show an unphysical divergence for $t > 25$, the Keldysh-OCCD results are stable. The two theories largely agree at short times where we would expect both approximations to be valid. However at longer times, different physics is predicted: Keldysh-CCSD yields a sharp drop in the valence to conduction population transfer at the point where there is a strong violation of total particle number conservation, while Keldysh-OCCD predicts that the valence to conduction population transfer continues at a reduced rate after the pulse ends. 

\subsection{Molecular H$_2$ in a laser field}

We further apply the Keldysh-OCCD theory to a field-driven molecular system. The purpose of this calculation is to provide a simple benchmark so that future implementations of Keldysh-OCCD and similar theories can be easily tested. The total molecular Hamiltonian is described in Ref.~\onlinecite{Huber2011}, where  
\begin{equation}
%\begin{split}
    H(t)=H(0)-\Vec{\mu}\cdot\Vec{A}(t). 
%\end{split}
\end{equation}
    $H(0)$ is the time-independent molecular Hamiltonian, and
\begin{equation}
%\begin{split}
    \Vec{\mu}=-\sum_i^N\Vec{r}_i+\sum_A^{N_A}Z_A\Vec{R}_A
%\end{split}
\end{equation}
is the molecular dipole operator for $N$ electrons and $N_A$ nuclei. We use a field of the same form as in Equation~\ref{pulse}. 

\begin{figure}[h]
    \centering
    \includegraphics[]{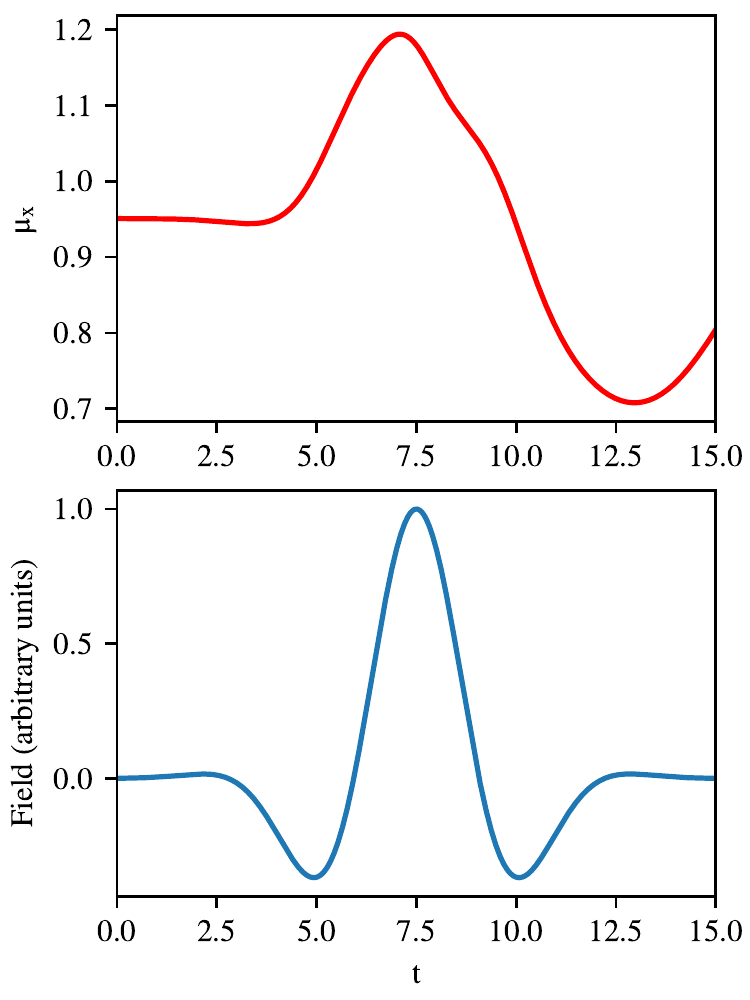}
    \caption{$\mu_x$ from Keldysh-OCCD as a function of time for a field-driven molecular H$_2$ benchmark. The shape of the electric field as a function of time is also plotted in the lower panel. }
    \label{fig:H2}
\end{figure}

In Figure \ref{fig:H2}, we show the $x$-component of the dipole moment for H${}_2$ in the STO-3G basis~\cite{sto3g} with the molecule aligned along the x-axis. We use field parameters $t_0=15.0$, $\sigma=2.0$, $\omega=1.0$, $\Vec{z}=(1.0,1.0,1.0)$ and $T=1.0$, $\mu=0$. The raw data for $\mu_x(t)$ are provided in the supporting information. 

\subsection{Single Impurity Anderson Model}

Finally we apply the Keldysh-OCCD method to  transport in the single impurity Anderson model, with a central impurity (``dot") coupled to two one-dimensional leads. We use a Hamiltonian of the form
\begin{equation}
    \hat{H}=\Hat{H}_{\text{dot}}+\Hat{H}_{\text{leads}}+\Hat{H}_{\text{dot-leads}}+\Hat{H}_{\text{bias}}
\end{equation}
where
%\begin{widetext}
\begin{align}
    \Hat{H}_{\text{dot}} &= V_gn_d+Un_{d\uparrow}n_{d\downarrow}\\
    \Hat{H}_{\text{leads}} &= -t_{\text{leads}}\sum_{p\sigma}(a_{Lp\sigma}^\dagger a_{Lp+1\sigma}+a_{Rp\sigma}^\dagger a_{Rp+1\sigma}+ \mathrm{h.c.}) \\
    \Hat{H}_{\text{dot-leads}} &= -t_{\text{hyb}}\sum_\sigma(a_{L1\sigma}^\dagger a_{d\sigma}+a_{R1\sigma}^\dagger a_{d\sigma}+\mathrm{h.c.}) \\
    \Hat{H}_{\text{bias}} &= \frac{V}{2}\sum_{p\sigma}(a_{Lp\sigma}^\dagger a_{Lp\sigma}-a_{Rp\sigma}^\dagger a_{Rp\sigma})
\end{align}
%\end{widetext}
as used previously in Ref.~\onlinecite{Kretchmer2018}. Here the impurity is associated with fermion operators $a^{(\dag)}_d$ and its Hamiltonian is parametrized by a gate voltage $V_g$ and Hubbard interaction $U$, while the left and right leads are associated with fermion operators $a^{(\dag)}_{Lp}$,  $a^{(\dag)}_{Rp}$, respectively, and are described by tight-binding Hamiltonians.
In the following calculations, the equilibrium state is generated by the Hamiltonian with parameters $t_{\text{leads}}=1.0$, $t_{\text{hyb}}=0.4$,  and zero bias $V=0$ for various specified temperatures and values of $U$. For all temperatures, 
the total system 
%is made of 16 sites and we adjust the chemical potential to obtain 16 
has the same number of particles as the number of sites with total $S_z=0$. A Hartree Fock calculation is performed at zero temperature, and these orbitals are used in the equilibrium CC calculations.

The dynamics is then generated by applying a small bias, $V=-0.005$, and the other parameters are kept fixed. The dynamics can be characterized by the time-dependent current across the dot, which we compute as the average of the current between the dot and its closest left and right neighbors $J(t)=(J_L(t)+J_R(t))/2$, where
\begin{equation}
    J_L(t)=%-2\pi\frac{ite}{h}
    -it_\text{hyb}\sum_\sigma\langle a_{L1\sigma}^\dagger a_{d\sigma}-a_{d\sigma}^\dagger a_{L1\sigma}\rangle
\end{equation}
\begin{equation}
    J_R(t)=%-2\pi\frac{ite}{h}
    -it_\text{hyb}\sum_\sigma\langle a_{d\sigma}^\dagger a_{R1\sigma}-a_{R1\sigma}^\dagger a_{d\sigma}\rangle
\end{equation}
and the bracket denotes the expectation value with respect to the Keldysh-OCCD density matrix. 
%Furthermore, we use atomic unit where $\hbar=1$ and $e=1$. 

\begin{figure}[h]
    \centering
    \includegraphics[]{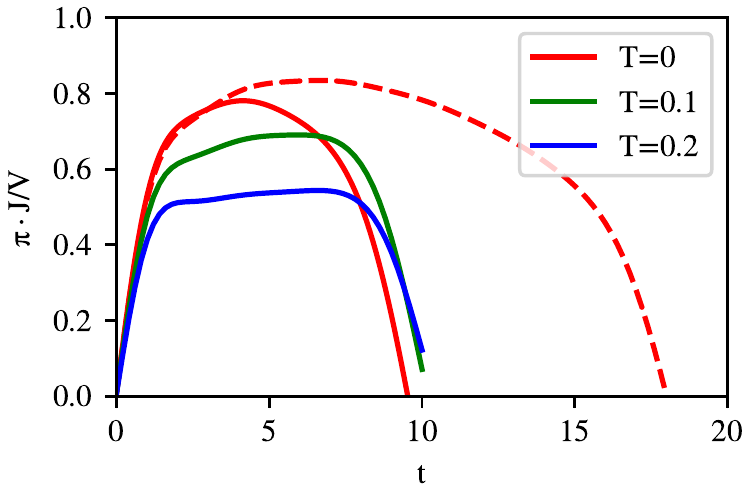}
    \caption{Time-dependent current at different temperatures for a gate of $V_g=-U/2$. The solid lines are systems with 16 sites and the dashed line is a system with 32 sites. }
    \label{fig:J}
\end{figure}

Figure~\ref{fig:J} shows an example of the time-dependent current divided by the bias for several different temperatures at $U=1.0$.
%using a total of 16 sites. 
As discussed, for example, in Ref.~\onlinecite{AlHassanieh2006}, the current will quickly reach its steady state in the infinite-size limit, but for finite leads, the finite system size produces an oscillatory behavior. In the case of 16 sites, we propagate for up to a time of $t=10.0$ corresponding to half the oscillation period, which is sufficient to extract the physics of the system pertaining to the infinite-size steady state. 

\begin{figure}[h]
    \centering
    \includegraphics[]{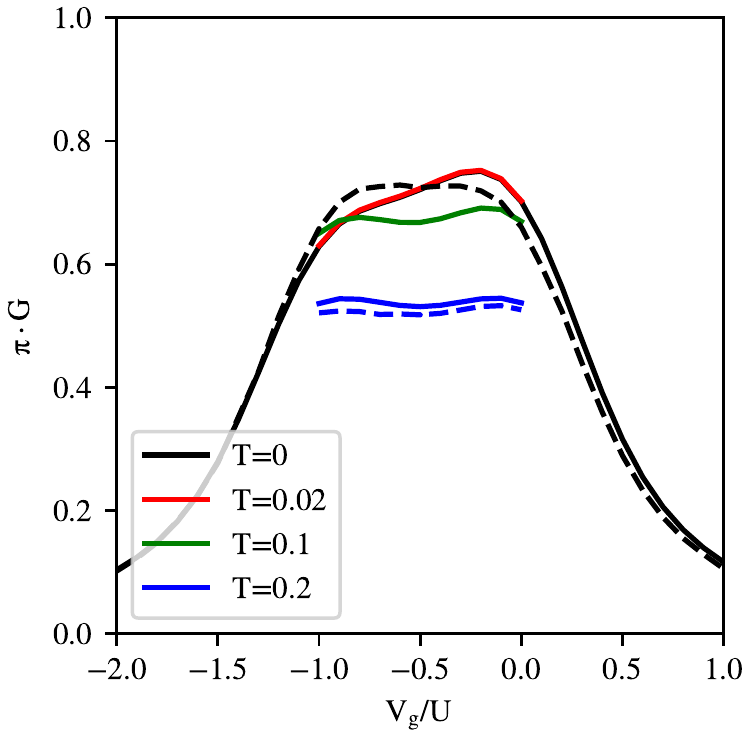}
    \caption{Conductance as a function of $V_g$ for various temperatures for 16 sites. The solid lines are Keldysh-OCCD results at the indicated temperatures and the dashed lines are the DMRG results at the corresponding temperatures.}
    \label{fig:Vg}
\end{figure}

Figure~\ref{fig:Vg} shows the conductance $G$ as a function of gate voltage $V_g$ for the 16 site model and $U=1.0$ at various temperatures  as computed from Keldysh-OCCD, as well as from reference density matrix renormalization group (DMRG) results. The conductance is computed as the  current divided by bias averaged over the plateau region from $t=2.0$ to $t=8.0$. The zero-temperature Keldysh-OCCD result is computed using a zero temperature implementation similar to that described in Sato \textit{et al}\cite{Sato2018CC}. The reference DMRG results at both zero- and finite-temperature are computed using a time-step targetting time-dependent DMRG method\cite{Feuguin2005A,Feuguin2005B,Ronca2017,Li2020} as implemented in the PyBlock3 software package\cite{block2,pyblock3} interfaced with the HPTT library\cite{hptt2017}.
The ``Kondo peak" in the low-temperature limit can be observed in the shape of a high conductance plateau, arising from many-body effects. As described elsewhere (see e.g. Ref.~\onlinecite{hewson1997kondo}), the Kondo resonance marks the increase of the density of states of the  impurity around the Fermi surface of the leads due to the spin interaction between a particle  near the Fermi-surface of the leads, and a particle on the impurity, and this resonance results in a high tunneling probability. The Kondo effect is only observed at temperatures below $T_K \sim e^{-1/j\rho}$ where $j$ is the (typically small) effective exchange coupling between a particle in a lead state and a particle on the impurity, and $\rho$ is the lead density of states at the Fermi-energy.
We find that the plateau at $T=0$ does not reach the predicted $G=1/\pi$ unitary value at $-U/2$. The deviation from the unitary limit has been seen in previous calculations and can be attributed to finite-size errors\cite{AlHassanieh2006} (for example, Figure 6(b) of Ref.~\onlinecite{Kretchmer2018} plots the conductance at $V_g=-U/2$ as a function of system size, demonstrating the convergence to the unitary value $G=1/\pi$ (or more precisely, for the unit used in that work, $2e^2/h$) as the system size approaches infinity). 

\begin{figure}[h]
    \centering
    \includegraphics[]{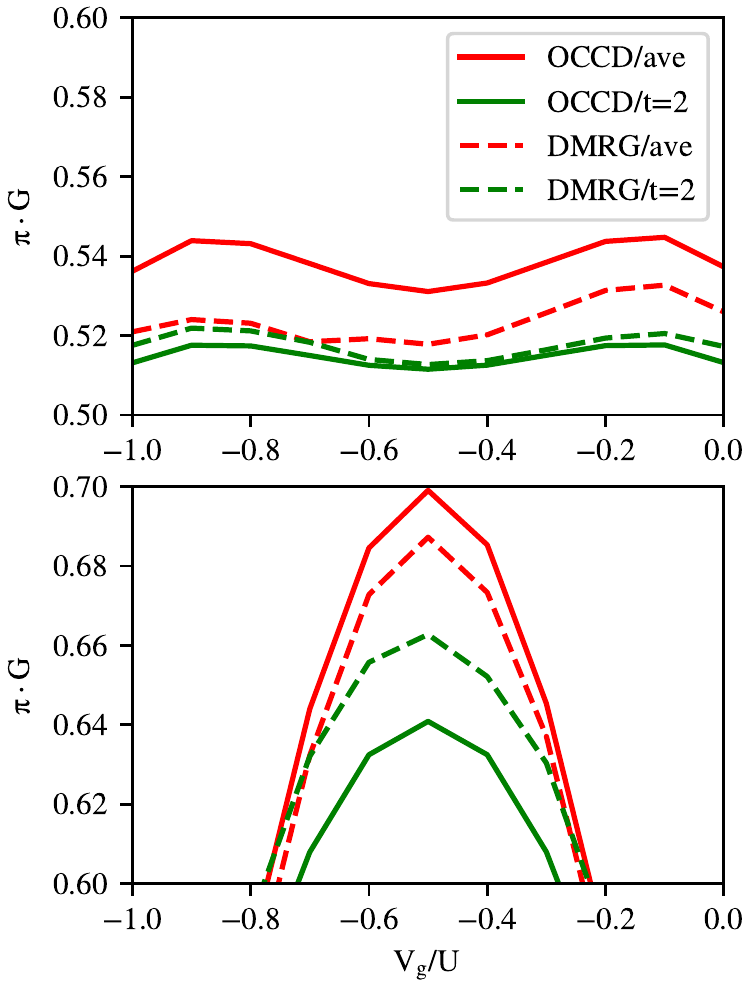}
    \caption{Conductance as a function of $V_g$ at $T=0.2$ for 16 sites. In both panels, red is the conductance computed as an average of $J/V$ in the window $t=2$ to $t=8$, and green is the conductance taken as the $J/V$ value at $t=2.0$. Solid lines are the Keldysh-OCCD results and dashed lines the DMRG results. The top panel corresponds to  an     interaction strength $U=1.0$. The lower panel shows test results with $U=0.0$ where Keldysh-OCCD is exact and DMRG is not due to  the finite bond-dimension. 
    }
    \label{fig:window}
\end{figure}

Figure~\ref{fig:window} shows a more detailed comparison of the Keldysh-OCCD results and DMRG results for the 16-site model at $T=0.2$ with  interaction $U=1.0$ in the top panel and vanishing interaction $U=0.0$ in the lower panel. For both the interacting and non-interacting case, the Keldysh-OCCD results are propagated with time step $dt=0.01$, while the DMRG results are obtained  using $dt=0.1$ and bond dimension $M=2000$ (we used a larger time-step in the DMRG to reduce the cost). We make note of two numerical aspects: (i) The conductance is very sensitive to the window over which the current is averaged. This can be seen from computing the conductance in two different ways, i.e.  as an average over the current divided by bias from $t=2.0$ to $t=8.0$ as shown in red, and as the current divided by the bias value at $t=2.0$ as shown in green. As shown in the top panel, different averaging windows result in both an overall vertical shift of the conductance, and a small change in the shape of the curve. (ii) For the non-interacting case in the lower panel, where the Keldysh-OCCD method is exact, there is still a difference between the Keldysh-OCCD conductance and the DMRG conductance, coming from the bond dimension truncation. 
%As pointed out in Ref.~\onlinecite{AlHassanieh2006}, DMRG still has an error depending on the truncation of bond dimension at $U=0$. 
Thus given the sensitivity of $G$ to the averaging window of the current, as well as the DMRG bond dimension truncation error, the Keldysh-OCCD and DMRG results in Figs.~\ref{fig:Vg} and~\ref{fig:window} are in very good agreement.
%for both the interacting and non-interacting cases. 

%changes not  a zoomed-in view of the same $T=0.2$ result in the top panel, where the Keldysh-OCCD conductance exhibits a valley at $V_g=-U/2$ in the high-temperature limit. The same feature is also seen in the DMRG conductance which is within {\color{red}5\%} of the Keldysh-OCCD conductance.
%This valley marks the vanishing of the Kondo peak and the onset of the Coulomb blockade regime, where the empty and occupied impurity states must be shifted by the gate voltage to align with the Fermi levels of the leads to observe conductance.

\begin{figure}[h]
    \centering
    \includegraphics[]{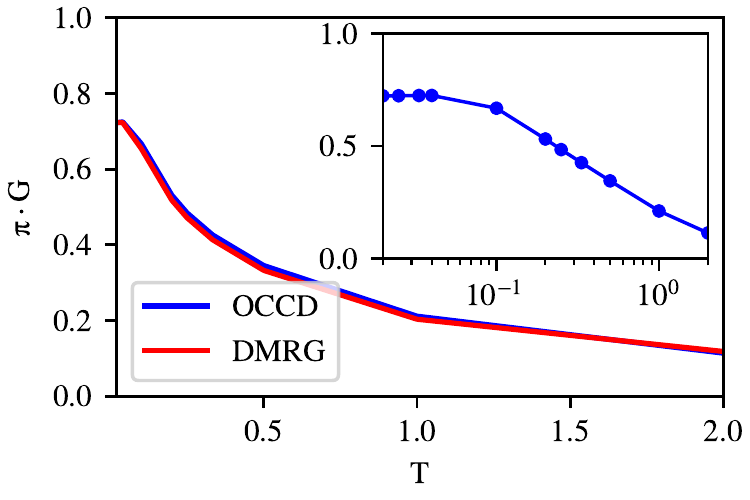}
    \caption{Conductance from Keldysh-OCCD and DMRG as a function of temperature at $V_g=-U/2$ for 16 sites, $U=1.0$. The inset shows the Keldysh-OCCD data with the temperature on a log scale.}
    \label{fig:T}
\end{figure}

The temperature-dependence of the conductance at $V_g=-U/2$ ($U=1.0$) is shown in Figure~\ref{fig:T}, where the inset plots the Keldysh-OCCD and DMRG results with data points marked and the temperature axis on a log-scale. Analytic treatments indicate that the peak conductance has a logarithmic dependence on temperature~\cite{hewson1997kondo} which is consistent with our data shown in the inset. 
%temperatu which shows the logarithmic dependence of conductance of temperature. {\color{red}How does it show this?} 
%As discussed in Ref.~\onlinecite{hewson1997kondo}, in a 1D system the Kondo effect increases the conductance in the low-temperature limit. This is because the lead energy levels are non-degenerate, and the increased spin interaction between the lead electron near the Fermi-level and dot electron increases tunneling probability of the lead electrons through the dot. On the other hand, for 3D systems, the lead energy levels are degenerate. The increased interaction between a lead electron near the Fermi-surface with the dot electron scatters other lead electrons near the Fermi-surface, which increases resistance. 
Thus our results show that at moderate interaction strength, Keldysh-OCCD is able to correctly capture the physics of the single impurity Anderson model, including the non-equilibrium Kondo physics and its temperature-dependence. 

\section{Conclusions}

In this work, we present a modification of the Keldysh coupled cluster theory that restores local and global conservation of one-particle quantities via an optimal orbital dynamics. On a variety of models and simple ab initio systems, we have demonstrated that such conservation laws are indeed obeyed within the Keldysh orbital-optimized coupled cluster doubles approximation (Keldysh-OCCD), with a concomitant improvement of the predicted dynamics, especially at longer times. In the single impurity Anderson model, we   qualitatively reproduce the temperature dependent transport physics, including that associated with the Kondo plateau. We believe this will be useful in the ab initio modeling of Kondo transport in the future.

However, there remain important challenges in the practical application of the Keldysh coupled cluster formalism: (i) at the doubles level, the cost and memory scaling 
%in applying the current Keldysh-OCCD method we still face several challenges which suggests future work: (i) As discussed in Section~\ref{sec:implementation}, the present Keldysh-OCCD method has the same cost and memory scaling as the previous Keldysh-CCSD. This high computational cost 
remains a barrier to many interesting applications. To decrease the computational cost, a potential future direction is to replace the optimal orbital dynamics
%rotation step  generated by the CCD 1- and 2-RDM 
by the orbital dynamics of time-dependent Hartree-Fock. This will not exactly preserve Ehrenfest's theorem for 1-particle dynamics as does the present Keldysh-OCCD approximation. However, for weakly interacting systems, the Hartree-Fock orbital update should still give better results than  Keldysh coupled cluster approximations without orbital dynamics.
%since the dominant 1-particle effect will be included in a unitary manner. 
(ii) At a formal level, the
fact that the state 
generated by the finite-temperature coupled cluster theory is not in general a stationary state of the dynamical theory leads to an ambiguity in the definition of the equilibrium state.

\begin{acknowledgements}
This work was supported by the US Department of Energy, Office of Science, via grant no. DE-SC0018140. Benchmarks generated by DMRG used PyBlock3, a code developed with support from the US National Science Foundation under grant no. CHE-2102505.
GKC thanks Emanuel Gull for discussions. GKC is a Simons Investigator in Physics and is part of the Simons Collaboration on the Many-Electron Problem.
\end{acknowledgements}
\appendix

\section{Orbital rotations and Ehrenfest's theorem at zero and finite temperature}\label{sec:AEhrenfest}

Here we show that Ehrenfest's theorem is restored for one-particle properties by including the optimal orbital dynamics into a zero-temperature time-dependent wavefunction ansatz. Let $\Psi(y_\nu,p)$ be the time-dependent wavefunction ansatz, where $y_\nu(t)$ are the variational parameters (e.g. such as the CI coefficients) and $p(t)$ is the orbital basis. The equation of motion of $y_\nu(t)$ and $p(t)$ can be determined from a time-dependent variational principle \cite{Sato2013,Sato2016,Sato2018CC,Sato2018CI,Kretchmer2018} by making the action,
\begin{equation}
    S[\Psi]=\int_0^Tdt\langle\Psi|H-i\partial_t|\Psi\rangle,
\end{equation}
stationary w.r.t small variations 
\begin{equation}
    |\delta\Psi\rangle=\delta y_\nu\partial_{y_\nu}|\Psi\rangle+\Delta|\Psi\rangle
\end{equation}
where the variation w.r.t. the orbitals, $\Delta|\Psi\rangle$, is parameterized by an anti-Hermitian 1-body operator,  $\Delta$\cite{Kretchmer2018}. Setting $\delta S=0$ leads to 
\begin{equation}\label{Aamp}
    0=\partial_{y_\nu^*}\langle\Psi|\left(H|\Psi\rangle-i|\dot{\Psi}\rangle\right)
\end{equation}
\begin{equation}\label{Aorb}
    0=i\langle\Psi|\Delta|\dot{\Psi}\rangle+i\langle\dot{\Psi}|\Delta|\Psi\rangle+\langle\Psi|[H,\Delta]|\Psi\rangle
\end{equation}
where solving equation (\ref{Aorb}) for each orbital pair is equivalent to enforcing Ehrenfest's theorem for the 1-particle density matrix elements, and hence for any 1-particle property.

Furthermore, the energy is conserved: the time-dependence of $\Psi$ can be expressed as \begin{equation}\label{Atd}
    |\dot{\Psi}\rangle=\dot{y}_\nu\partial_{y_\nu}|\Psi\rangle+X|\Psi\rangle
\end{equation}
where $X$ is anti-Hermitian and parameterizes the time-dependence of the orbitals\cite{Kretchmer2018}. Then the energy derivative
\begin{align}
    \frac{d}{dt}\langle H\rangle
    &=\langle\dot{\Psi}|H|\Psi\rangle+\langle\Psi|H|\dot{\Psi}
    \rangle \nonumber\\
    &=\left(\dot{y}_\nu\partial_{y_\nu^*}\langle\Psi|\right)H|\Psi
    \rangle - \langle\Psi|XH|\Psi\rangle \nonumber\\
    &+\langle\Psi|H\left(\partial_{y_\nu}|\Psi\rangle\dot{y}_\nu
    \right) + \langle\Psi|HX|\Psi\rangle \nonumber\\
    &=i\left(\dot{y}_\nu\partial_{y_\nu^*}\langle\Psi|\right)|
    \dot{\Psi}\rangle - 
    i\langle\dot{\Psi}|\left(\partial_{y_\nu}|\Psi\rangle\dot{y}_\nu
    \right)\nonumber\\
    &-i\langle\Psi|X|\dot{\Psi}\rangle
    -i\langle\dot{\Psi}|X|\Psi\rangle\nonumber\\
    &=i\left(\langle\dot{\Psi}|\right)|\dot{\Psi}\rangle - 
    i\langle\dot{\Psi}|\left(|\dot{\Psi}\rangle\right)
=0
\end{align}
where we use equation \eqref{Aamp} and \eqref{Aorb} for the 3rd equality, and equation \eqref{Atd} for the 4th equality. 

At finite temperature, the idea is analogous: orbital optimization based on an action principle can be used to satisfy Ehrenfest's theorem. %\sout{To show how this works explicitly, we will demonstrate it for the case of Keldysh orbital-optimized second-order perturbation theory (Keldysh OMP2), as the algebra easily generalizes to the case of Keldysh OCCD.}

We first derive the conditions satisfied by the stationary orbital dynamics. The Lagrangian in equations~\ref{full_Q} and ~\ref{theory1_L} can be written as
\begin{equation}
\begin{split}
\mathcal{Q}
&=\frac{1}{2}(\mathcal{L}+\mathcal{L}^{\ast})+\frac{i}{\beta}\int_Cdt E^{(1)}+\Omega^{(0)}
\end{split}
\end{equation}
where
\begin{equation}\label{E1}
E^{(1)}
=h_{ii}+R_{ii}+\frac{1}{2}\langle ij||ij\rangle=\langle H-iX\rangle_0
\end{equation}
and 
\begin{align}
\mathcal{L}
&=\frac{i}{\beta}\int_Cdt(E(t)+\tilde{\lambda}^\nu(t) S_\nu(t))\\
&+\frac{i}{\beta}\int_Cdt\tilde{\lambda}^\nu(t)(-i\partial_ts_\nu(t))\\
&=\frac{i}{\beta}\int_Cdt\langle H-iX\rangle_{\text{CC}_N}\label{expectationCC}\\
&+\frac{i}{\beta}\int_Cdt\tilde{\lambda}^\nu(t)(-i\partial_ts_\nu(t))\label{ddot}
\end{align}
where the term $-iX$ comes from the modification of the integrals as in equation~\ref{eqn:formal_R}. We use the notation $\langle...\rangle_0$ to denote an expectation value with respect to the mean-field 1-RDM $p_{pq}$ in equation~\ref{rdm0} and 2-RDM $p^{pq}_{rs}=p_{pr}p_{qs}-p_{ps}p_{qr}$, and $\langle...\rangle_{\text{CC}_N}$ to  denote an expectation value with respect to the coupled-cluster normal-ordered 1- and 2-RDMs defined in equations~\ref{rdmCC1}-\ref{rdmCC2}. 
As in the zero-temperature case, the orbital variation can be parameterized by an antihermitian matrix $\Delta$
\begin{equation}
\delta C_{pq}=C_{pr}\Delta_{rq}
\end{equation}
and then the variation of the modified Hamiltonian is
\begin{equation}
\begin{split}
\delta (h_{pq}-iX_{pq})=(h_{pr}-iX_{pr})\Delta_{rq}-\Delta_{pr}(h_{rq}-iX_{rp})
\end{split}
\end{equation}
\begin{equation}
\begin{split}
\delta\langle pq||rs\rangle
&=\langle pq||xs\rangle\Delta_{xr}+\langle pq||rx\rangle\Delta_{xs}\\
&-\Delta_{px}\langle xq||rs\rangle-\Delta_{qx}\langle px||rs\rangle
\end{split}
\end{equation}
 which can be compactly denoted as $[H-iX,\Delta]$. In the orbital variation of the action $\mathcal{Q}$, the variation of terms~\ref{E1},~\ref{expectationCC} comes from the variation of the modified Hamiltonian tensors
\begin{equation}
\begin{split}
&\delta E^{(1)}
=\langle[H-iX,\Delta]\rangle_0\\
&\delta(E(t)+\tilde{\lambda^\nu(t)S_\nu(t)})
=\langle[H-iX,\Delta]\rangle_{\text{CC}_N},
\end{split}
\end{equation}
and the variation of term~\ref{ddot} gives
\begin{equation}
\begin{split}
-i(\delta\tilde{\lambda}^\nu)\dot{s}_\nu+i\dot{\tilde{\lambda}}^\nu\delta s_\nu=-i \mathrm{Tr}(\dot{d}\Delta)
\end{split}
\end{equation}
where $\dot{d}$ is the time-derivative of the CC 1-RDM in Equation~\ref{rdm1}. Thus the orbital gradient of $\mathcal{Q}$ is
\begin{equation}
\delta\mathcal{Q}=\langle [H,\Delta]\rangle_\text{CC}-i\langle[X,\Delta]\rangle_\text{CC} -i\mathrm{Tr}(\dot{d}\Delta)
\end{equation}
where $\langle...\rangle_\text{CC}$ denotes the expectation value computed using the CC 1- and 2-RDMs given in  equations~\ref{rdm1},~\ref{rdm2}. Setting $\Delta\mathcal{Q}=0$ for each orbital pair $\Delta_{uv}$ gives

%{\color{red}Insert Theory 1, then connect to Ehrenfest's discussion below. Also make sure notation is properly defined, e.g. $[v, \Delta]$ needs to be defined explicitly. Then after this, at the very end I will insert a few words about theory 2/theory 3 leading to non-zero oo/vv rotations.}

%that this is true for the Keldysh OCCD theory presented here, we will first write Ehrenfest's theorem (Equation~\ref{ehrenfest}) in terms of densities in the MO basis:
%\begin{equation*}
%	\sum_{uv}
%	A_{uv}\dot{d}_{vu} + \sum_{uvq}d_{qu}A_{uv}X_{vq} - \sum_{uvp}d_{vp}A_{uv}X_{pu}
%\end{equation*}
%\begin{equation}
%	\qquad = i\sum_{uv}\left[\mathcal{F}_{vu} - \mathcal{F}_{uv}^{\ast}\right]A_{uv}.
%\end{equation}
% Since this equation should hold for any operator, it must be separately satisfied for each $uv$ pair:
\begin{equation}\label{eqn:EhrenfestDen}
	\dot{d}_{vu} + \sum_{q}d_{qu}X_{vq} - \sum_{p}d_{vp}X_{pu}
	= i\left[\mathcal{F}_{vu} - \mathcal{F}_{uv}^{\ast}\right].
\end{equation}
where 
%the $X$ matrix is the time derivative of the orbitals defined in Equation~\ref{eqn:CDeriv}, and
the $\mathcal{F}$ matrix is defined in equation~\ref{eqn:orbF}.
For Keldysh OCCD, only the ``occupied-occupied" and ``virtual-virtual" blocks of the 1-RDM are non-zero. This means that the only non-zero blocks of Equation~\ref{eqn:EhrenfestDen} are
\begin{align}
	\dot{d}_{ij} + \sum_{k}d_{kj}X_{ik} - \sum_{k}d_{ik}X_{kj} &= i\left[\mathcal{F}_{ij} - 
	\mathcal{F}_{ji}^{\ast}\right] \label{eqn:oo}\\
	\dot{d}_{ab} + \sum_{c}d_{cb}X_{ac} - \sum_{c}d_{ac}X_{cb} &= i\left[\mathcal{F}_{ab} - 
	\mathcal{F}_{ba}^{\ast}\right] \label{eqn:vv}\\
	\sum_{b}d_{ba}X_{ib} - \sum_{j}d_{ij}X_{ja} &= i\left[\mathcal{F}_{ia} - 
	\mathcal{F}_{ai}^{\ast}\right] \\
	\sum_{j}d_{ji}X_{aj} - \sum_{b}d_{ab}X_{bi} &= i\left[\mathcal{F}_{ai} - 
	\mathcal{F}_{ia}^{\ast}\right].
\end{align}
For an anti-hermitian matrix $X$, the final two equations are complex conjugates of each other and we only need to satisfy one of them, which we rewrite as
\begin{equation}
	\sum_{j}d_{ji}R_{aj} - \sum_{b}d_{ab}R_{bi} = \left[\mathcal{F}_{ai} - 
	\mathcal{F}_{ia}^{\ast}\right]
\end{equation}
where we have defined $R \equiv -iX$. This is  the 
precise form of the orbital equation previously shown in Equation~\ref{eqn:ft_orb}
that comes from stationarity of the Lagrangian. Furthermore, it can be shown that equations~\ref{eqn:oo}, ~\ref{eqn:vv} trivially hold for any occupied-occupied and virtual-virtual rotation, and hence those rotations can be set to zero. 

We can write Ehrenfest's theorem in a rotating basis  at finite temperature in terms of the reduced density matrices and matrix elements of the operator $A$ in the orbitals defined in Equation~\ref{eqn:CDeriv}, 
\begin{equation*}
	\sum_{uv}
	A_{uv}\dot{d}_{vu} + \sum_{uvq}d_{qu}A_{uv}X_{vq} - \sum_{uvp}d_{vp}A_{uv}X_{pu}
\end{equation*}
\begin{equation}
	\qquad = i\sum_{uv}\left[\mathcal{F}_{vu} - \mathcal{F}_{uv}^{\ast}\right]A_{uv}.
\end{equation}
%The $X$ matrix is the time derivative of the orbitals defined in Equation~\ref{eqn:CDeriv}, and the $\mathcal{F}$ matrix is defined in Equation~\ref{eqn:orbF}. 
Since this equation should hold for any operator, it must be separately satisfied for each $uv$ pair. This immediately yields equation~\ref{eqn:EhrenfestDen}, where the reduced density matrices correspond to their definition in coupled cluster theory.  Thus, we see that the stationarity of the coupled cluster Lagrangian with respect to orbital variations leads to the satisfaction of Ehrenfest's theorem.

\section{Conserving approximations}
\label{sec:conserving}

There are several equivalent ways to define a conserving approximation in the theory of Green's functions. To draw a close correspondence with the approach in this work, we define a conserving approximation to be one where the Green's function on the Keldysh contour $G_{pq}(t_1,t_2) = i \mathcal{T}_C \langle a_p(t_1) a^\dag_q(t_2) \rangle$ (where $\mathcal{T}_c$ indicates contour time-ordering) makes the Luttinger-Ward functional $A$ stationary, defined as
\begin{align}
A[\underline{G}] = -\frac{1}{\beta}  \mathrm{Tr} [\log (-\underline{G}_0^{-1} + \underline{\Sigma}) + \underline{\Sigma}\underline{G}] + \Phi[\underline{G}] \end{align}
where $\underline{\Sigma}[\underline{G}]$ is the self-energy, $\underline{G}_0$ is the zeroth order Green's function, the underline notation indicates that the Green's function and self-energy elements $G_{pq}$, $\Sigma_{pq}$ are themselves $2\times 2$ matrices, with row/columns labelling pairs of contour indices along the forwards and backwards contours, and $\mathrm{Tr}$ integrates over contour time as well as sums over the contour and orbital indices. $\Phi[\underline{G}]$ is a sum of closed diagrams of $\underline{G}$ and the two-particle interaction, and $\underline{\Sigma}[\underline{G}]_{qp}(t_2,t_1) = \mathrm{sgn} \cdot \delta \Phi[G]/\delta \underline{G_{pq}}(t_1,t_2)$ where $\mathrm{sgn}$ introduces the appropriate sign for different pairs of contour indices in the elements of $\underline{\Sigma}$. For a more detailed explanation of the terminology, see e.g. Ref.~\cite{kita2010introduction}. We neglect some subtleties related to convergence on the real-time contour discussed in Ref.~\cite{hofmann2013nonequilibrium}. For an equilibrium problem, the Luttinger-Ward functional evaluates to the thermodynamic grand potential, thus it is an analog of the finite-temperature and Keldysh coupled cluster Lagrangians described here.

As we have argued in the main text, Ehrenfest's theorem arises when the dynamics is stationary under orbital variations.
The local conservation laws for one-particle quantities, such as the density and momentum density, are a consequence of Ehrenfest's theorem for one-particle quantities. In the case of the coupled cluster Lagrangian, variations with respect to $T$ and $\Lambda$ (i.e. changing the values of the amplitudes) do not completely capture the space of variations when the underlying orbitals are changed. (In other words, even when the coupled cluster Lagrangian is stationary w.r.t. $T, \Lambda$, under an orbital rotation that changes $H \to e^{i\epsilon R} H e^{-i\epsilon R}$, there is not a small change in the values of the amplitudes which completely cancels this rotation). However, in the case of the Luttinger-Ward functional, stationarity with respect to the Green's function implies stationarity with respect to the underlying orbitals, because a small change in $H \to e^{i\epsilon R} H e^{-i\epsilon R}$ can be cancelled by a corresponding small change in the Green's function (with a small abuse of notation,  $\underline{G} \to e^{-i\epsilon R} \underline{G} e^{i\epsilon R}$) since all quantities in the action correspond to closed diagrams of $H$ and $\underline{G}$. 

\section{Stationarity of perturbation theory}\label{app:stationary}

Here we briefly show that finite-temperature time-dependent perturbation theory yields stationary equilibrium 
observables. Consider a Hamiltonian $H(\lambda)=h+\lambda V$ where $h$ is the zeroth order piece.
The time-dependent observable $O$ and its equilibrium value are identical under propagation by the equilibrium Hamiltonian since
\begin{align}
    & Z^{-1} \mathrm{tr} \ e^{-\beta H(\lambda)} e^{iH(\lambda) T} O e^{-iH(\lambda) T} \notag\\
    & = Z^{-1} \mathrm{tr} \ e^{-iH(\lambda) T} e^{-\beta H(\lambda)} e^{iH(\lambda) T} O \notag\\
    &= Z^{-1} \mathrm{tr} \ e^{-\beta H(\lambda)} e^{-iH(\lambda)T} e^{iH(\lambda T)} O \notag\\
    &= Z^{-1} \mathrm{tr} \ e^{-\beta H(\lambda)} O
\end{align}
where $Z$ is the partition function, the second line follows from cyclic invariance, and the third line from commuting operators, which does not require the trace. The above is an identity which holds for all $\lambda$, therefore it is true order by order in $\lambda$, and that means that the perturbation expansion of the left hand side and right hand side must agree. The need to include all time-orderings to obtain stationarity in an approximate theory is because the above result relies on commuting the imaginary and real-time propagations past each other, which is equivalent to changing the time-ordering of interactions on those branches.

\section{Coupled cluster equations}\label{app:equations}
The kernels which precisely determine the Keldysh-OCCD method closely resemble the zero-temperature OCCD equations. The E kernel
is given by 
\begin{equation}
    \mathrm{E}(t) = \frac{1}{4}\sum_{ijab}\bra{ij}\ket{ab}s_{ij}^{ab}(t).
\end{equation}
The S and L kernels which determine the equations of motion for the $s$ and $\tilde{\lambda}$ amplitudes respectively are given by
\begin{widetext}
\begin{align}\label{eqn:ccS2}
	\text{S}_{ij}^{ab}(t) &= \bra{ab}\ket{ij}
	+ P(ab)\sum_{c}f_{bc}s_{ij}^{ac}(t) - 
	P(ij)\sum_{k}f_{kj} s_{ik}^{ab}(t)
	+ \frac{1}{2}\sum_{cd}\bra{ab}\ket{cd}s_{ij}^{cd}(t)
	\nonumber \\ &+\frac{1}{2}\sum_{kl}\bra{kl}\ket{ij}s_{kl}^{ab}(t) + P(ij)P(ab)\sum_{kc}\bra{kb}\ket{cj}s_{ik}^{ac}(t) 
	\nonumber \\
	&+ \frac{1}{4}\sum_{klcd} \bra{kl}\ket{cd}
	s_{ij}^{cd}(t)s_{kl}^{ab}(t) + \frac{1}{2}P(ij)P(ab)\sum_{klcd}\bra{kl}\ket{cd}
	s_{ik}^{ac}(t)s_{lj}^{db}(t)\nonumber \\ &
	- \frac{1}{2}P(ab)\sum_{klcd} \bra{kl}\ket{cd}s_{kl}^{ca}(t)s_{ij}^{db}(t)
	- \frac{1}{2}P(ij)\sum_{klcd}\bra{kl}\ket{cd}s_{ki}^{cd}(t)s_{lj}^{ab}(t)
\end{align}
\begin{align}\label{eqn:ccL2}
	\text{L}_{ab}^{ij}(t) &= \bra{ij}\ket{ab}\nonumber \\ 
	&+ P(ab)\sum_c \tilde{\lambda}^{ij}_{ac}(t)f_{cb} 
	- P(ij)\sum_k\tilde{\lambda}^{ik}_{ab}(t)f_{jk} 
	+ \frac{1}{2}\sum_{cd} \tilde{\lambda}^{ij}_{cd}(t) \bra{cd}\ket{ab} \nonumber \\
	&+\frac{1}{2}\sum_{kl}\tilde{\lambda}^{kl}_{ab}(t)\bra{ij}\ket{kl}
	+ P(ij)P(ab)\sum_{kc}\tilde{\lambda}^{ik}_{ac}(t)\bra{cj}\ket{kb}\nonumber \\ 
	&- P(ij)\frac{1}{2}\sum_{klcd}\tilde{\lambda}^{ik}_{ab}(t)
	\bra{jl}\ket{cd}s_{kl}^{cd}(t) 
	- P(ab)\frac{1}{2}\sum_{klcd}\tilde{\lambda}^{ij}_{ac}(t)
	\bra{kl}\ket{bd}s_{kl}^{cd}(t)\nonumber \\ 
	&+ P(ij)P(ab)\sum_{klcd}\tilde{\lambda}^{ik}_{ac}(t)
	\bra{lj}\ket{db}s_{kl}^{cd}(t)
	- P(ab)\frac{1}{2}\sum_{klcd}\tilde{\lambda}^{kl}_{ca}(t)\bra{ij}\ket{db}s_{kl}^{cd}(t)\nonumber \\
	&- P(ij)\frac{1}{2}\sum_{klcd}\tilde{\lambda}^{ki}_{cd}(t)\bra{lj}\ket{ab}s_{kl}^{cd}(t)
	+ \frac{1}{4}\sum_{klcd}\tilde{\lambda}^{kl}_{ab}(t)\bra{ij}\ket{cd}s_{kl}^{cd}(t)
	\nonumber \\
	&+ \frac{1}{4}\sum_{klcd}\tilde{\lambda}^{ij}_{cd}(t)\bra{kl}\ket{ab}s_{kl}^{cd}(t)
\end{align}
\end{widetext}
The density matrices which appear in the orbital equation (Equation~\ref{eqn:orbF}) are also used to compute properties. They can be obtained from the derivative of the Lagrangian with respect to the potential:
\begin{equation}\label{rdm1}
    d_{pq} = \frac{1}{2}\left[(d_N)_{pq} + (d_N)_{qp}^{\ast}\right] + p_{pq}.
\end{equation}
%\begin{widetext}
%\begin{align}
%d^{pq}_{rs} & = \frac{1}{2}\left[(d_N)^{pq}_{rs}+(d_N)^{rs}_{pq}\right]\\
%&+p_{pr}d_{qs}+p_{qs}d_{pr}-p_{ps}d_{qr}-p_{qr}d_{ps}\\
%&-p_{pr}p_{qs}+p_{ps}p_{qr}
%\end{align}
%\end{widetext}
\begin{equation}\label{rdm2}
\begin{split}
    d^{pq}_{rs} & = \frac{1}{2}\left[(d_N)^{pq}_{rs}+(d_N)^{rs}_{pq}{}^{\ast}\right]\\
&+p_{pr}d_{qs}+p_{qs}d_{pr}-p_{ps}d_{qr}-p_{qr}d_{ps}\\
&-p_{pr}p_{qs}+p_{ps}p_{qr}
\end{split}
\end{equation}
We have used $p$ to indicate the mean-field density matrix,
\begin{equation}\label{rdm0}
    p_{ij} = \delta_{ij}, \quad p_{ia} = p_{ai} = p_{ba} = 0,
\end{equation}
and $d_N$ for the coupled cluster contributions
\begin{align}
    (d_N)_{ia} &= 0\label{rdmCC1}\\
    (d_N)_{ba} &= \frac{1}{2}\sum_{ikc}\tilde{\lambda}_{cb}^{ki}(t)
    s_{ki}^{ca}(t)\\
    (d_N)_{ji} &= -\frac{1}{2}\sum_{akc}\tilde{\lambda}_{ca}^{kj}(t)
    s_{ki}^{ca}(t)\\
    (d_N)_{ai} &= 0
\end{align}
\begin{align}
    (d_N)^{ij}_{ab} & =\tilde{\lambda}^{ij}_{ab}(t)\\
    (d_N)^{aj}_{ib} & =\sum_{ck}s^{ac}_{ik}(t)\tilde{\lambda}^{jk}_{bc}(t)\\
    (d_N)^{cd}_{ab} & = \frac{1}{2}\sum_{ij}s^{cd}_{ij}(t)\tilde{\lambda}^{ij}_{ab}(t)\\
    (d_N)^{kl}_{ij} & = \frac{1}{2}\sum_{ab}s^{ab}_{ij}(t)\tilde{\lambda}^{kl}_{ab}\\
    (d_N)^{ab}_{ij} & = s^{ab}_{ij}(t)
    +P(ij)P(ab)\frac{1}{2}\sum_{klcd}s^{ac}_{ik}\tilde{\lambda}^{kl}_{cd}(t)s^{bd}_{jl}(t)\nonumber\\
    &-P(ab)\frac{1}{2}\sum_{klcd}s^{ad}_{kl}(t)\tilde{\lambda}^{kl}_{cd}(t)s^{cd}_{ij}(t)\nonumber\\
    &-P(ij)\frac{1}{2}\sum_{klcd}s^{cd}_{il}(t)\tilde{\lambda}^{kl}_{cd}(t)s^{ab}_{kl}(t)\nonumber\\
    &+\frac{1}{4}\sum_{klcd}s^{ab}_{kl}(t)\tilde{\lambda}^{kl}_{cd}(t)s^{cd}_{ij}(t)\label{rdmCC2}
\end{align}
Here and throughout this work we have assumed that the matrix elements include factors of the square root of the occupation numbers as in Equation~\ref{eqn:integrals}.

\bibliography{export}
\end{document}